\documentclass{interact}
\usepackage[latin1]{inputenc}
\usepackage{amsmath}
\usepackage{amsfonts}
\usepackage[english]{babel}
\usepackage{cite}
\usepackage{subfigure}
\usepackage{rotating}
\usepackage{color}
\usepackage{graphicx}
\usepackage{epstopdf}
\usepackage{graphicx}
\usepackage{rotating}

\usepackage{amssymb}
\usepackage{url}

\makeatletter
\renewcommand\@biblabel[1]{}
\renewenvironment{thebibliography}[1]
     {\section*{\refname}%
      \@mkboth{\MakeUppercase\refname}{\MakeUppercase\refname}%
      \list{}%
           {\leftmargin0pt
            \@openbib@code
            \usecounter{enumiv}}%
      \sloppy
      \clubpenalty4000
      \@clubpenalty \clubpenalty
      \widowpenalty4000%
      \sfcode`\.\@m}
     {\def\@noitemerr
       {\@latex@warning{Empty `thebibliography' environment}}%
      \endlist}
\makeatother

\begin{document}


\title{Public transportation in Great Britain viewed as a complex network}

\author{
\name{Robin de Regt\textsuperscript{a,c}\thanks{CONTACT Robin de Regt. Email: deregtr@uni.coventry.ac.uk}, Christian von Ferber\textsuperscript{a,c}, Yurij Holovatch\textsuperscript{b,a,c} and Mykola Lebovka\textsuperscript{d,e}}
\affil{\textsuperscript{a}Applied Mathematics Research Centre, Coventry University, Coventry, CV1 5FB, UK; \textsuperscript{b}Institute for Condensed Matter Physics, National Acad. Sci. of Ukraine,UA-79011 Lviv, Ukraine; \textsuperscript{c}$L^4$ Collaboration \& Doctoral College for the
Statistical Physics of Complex Systems, Leipzig-Lorraine-Lviv-Coventry; \textsuperscript{d}F.D. Ovcharenko Institute of Biocolloidal Chemistry, National Acad. Sci. of Ukraine, 03142 Kyiv, Ukraine;\textsuperscript{e} Sorbonne Universités, Université de Technologie de Compiègne, EA 4297, Centre de Recherches de Royallieu, BP 20529-60205 Compiègne Cedex, France}
}

\maketitle

\begin{abstract}
In this paper we investigate the topological and spatial features of public transport networks (PTN) within Great Britain. Networks investigated include London, Manchester, West Midlands, Bristol, national rail and coach networks during $2011$. Using methods in complex network theory and statistical physics we are able to discriminate PTN with respect to their stability; which is the first of this kind for national networks. Taking advantage of various fractal properties we gain useful insights into the serviceable area of stations. Moreover, we investigate universal load dynamics of these systems. These features can be employed as key performance indicators in aid of further developing efficient and stable PTN.
\end{abstract}


\section{Introduction}\label{0}

Over the last few decades society has become increasingly dependent on public
transport to facilitate commuters and the movement of commodities on both local and global scales. With transport having such a significant role in the economy of cities and countries it is becoming increasingly important to develop cost effective methods to evaluate the efficiency and robustness of existing public transport networks.
\\ 
One approach to study these networks is offered through complex network science, a recently established research field with a firm theoretical background and a broad range of applications. It has successfully explained numerous phenomena that 
have emerged in natural and man made systems involving separate agents 
connected via various types of interactions (Albert and Barab\'asi, 2002; Dorogovtsev and Mendes, 2003; Barrat et al., 2008; Newman, 2010).
\\
Very often underlying networks do not have direct geometrical interpretations (Guimera, 2007), for example in social networks that involve collaboration, acquaintances and friends. Here, one quantifies the network in terms of its topological features: node
degree distribution, connectivity, clustering, as will be discussed in more detail below and further elaborated on in the appendix.
\\
There are however networks which are shaped by 
their embedding in geometric space (Barth\'elemy, 2011), like transport networks for example. In addition to their topological properties, these networks are naturally quantified in terms of their geometric features. The latter being primarily defined by spatial coordinates of nodes and include Euclidean distances between nodes. 
\\
The purpose of this paper is to investigate the properties of PTN in Great Britain (GB) using both complex networks (i.e. topological) as well as spatial descriptions in order to gain useful insights into robustness and efficiency of PTN.
Currently, there exists a relatively broad literature on the application of a complex networks for public transport analysis (see, in particular, the discussion below).
Here, we add to the existing analysis and further explore certain topological measures which can be used to classify PTN with respect to their stability to random failures.
We add to this analysis by studying the fractals properties of these systems which offer insight into the serviceability and efficiency of PTN. 
Together both the topological and spatial features studied here may contribute to a better understanding of the underlying mechanisms governing PTN growth and modeling. 
This ignores important concepts of transportation, passenger flow and service frequency being beneath the major ones. In an attempt to avoid such types of flaws, we 
complete our study by analysing some features of PTN dynamics and network load.
\\
Another objective of our paper is to attract the interest of academics and
practitioners dealing with public transportation networks in furthering
the applications of the methods discussed here. So far, application of
these methods to transportation networks has been extensively
discussed and thoroughly approved on the pages of specialised
physical and complex system journals see e.g. the list of references at the beginning of the next section. Addressing this paper 
to the journal devoted to transportation we pursue an aim to interest its readers in more intensive practical application of the matters discussed here.
\\ 
The paper is organised as follows. In section
\ref{I} we present a brief review of the literature devoted to PTN
topological and spatial analysis.
Section \ref{II} describes the database we
use. The main results of our analysis are presented in section
\ref{IV}, where we discuss topological and spatial
aspects of several PTN: those of Greater London, Greater Manchester,
West Midlands, Bristol, and the national rail and coach networks of GB. Conclusions and an outlook are given in section \ref{V}.

\section{Review}\label{I}
\subsection*{Topology}
\begin{figure}
\centerline{\includegraphics[width=0.93\textwidth]{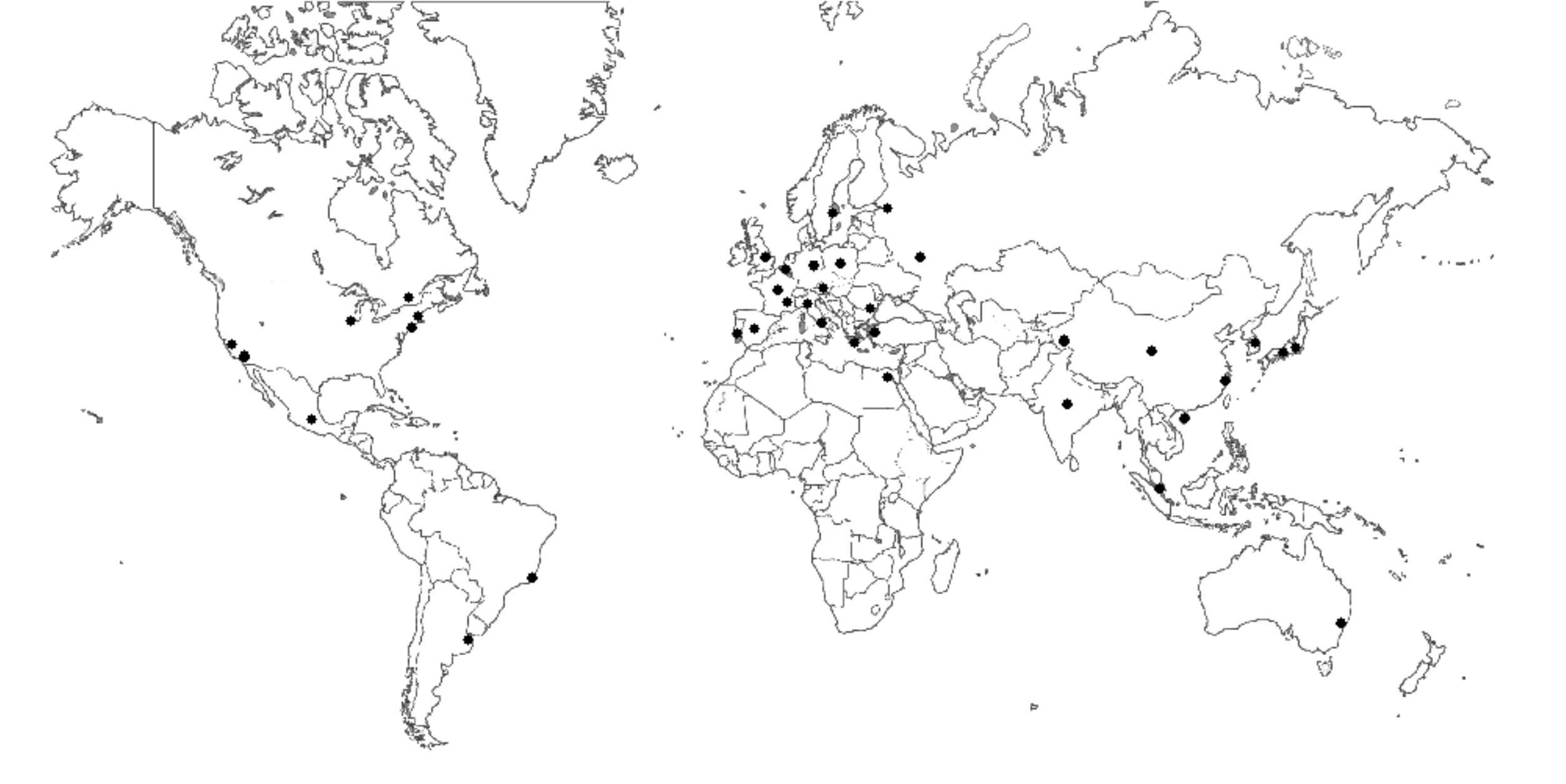}}
\caption{Map indicating various locations of the world where city PTN
have been analysed within the complex network science framework.
Topological characteristics of some of them are further displayed in
Table \ref{tab2}.}\label{fig1}
\end{figure}

\begin{figure}[ht]
\includegraphics[width=0.4\textwidth]{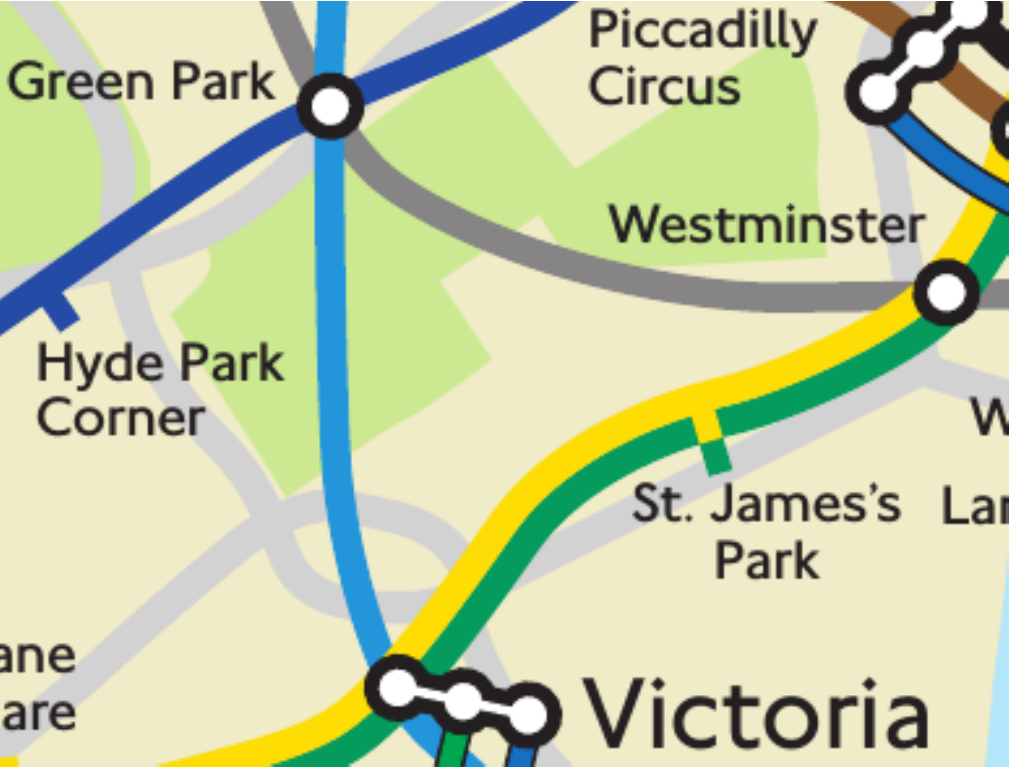}
\includegraphics[width=0.4\textwidth]{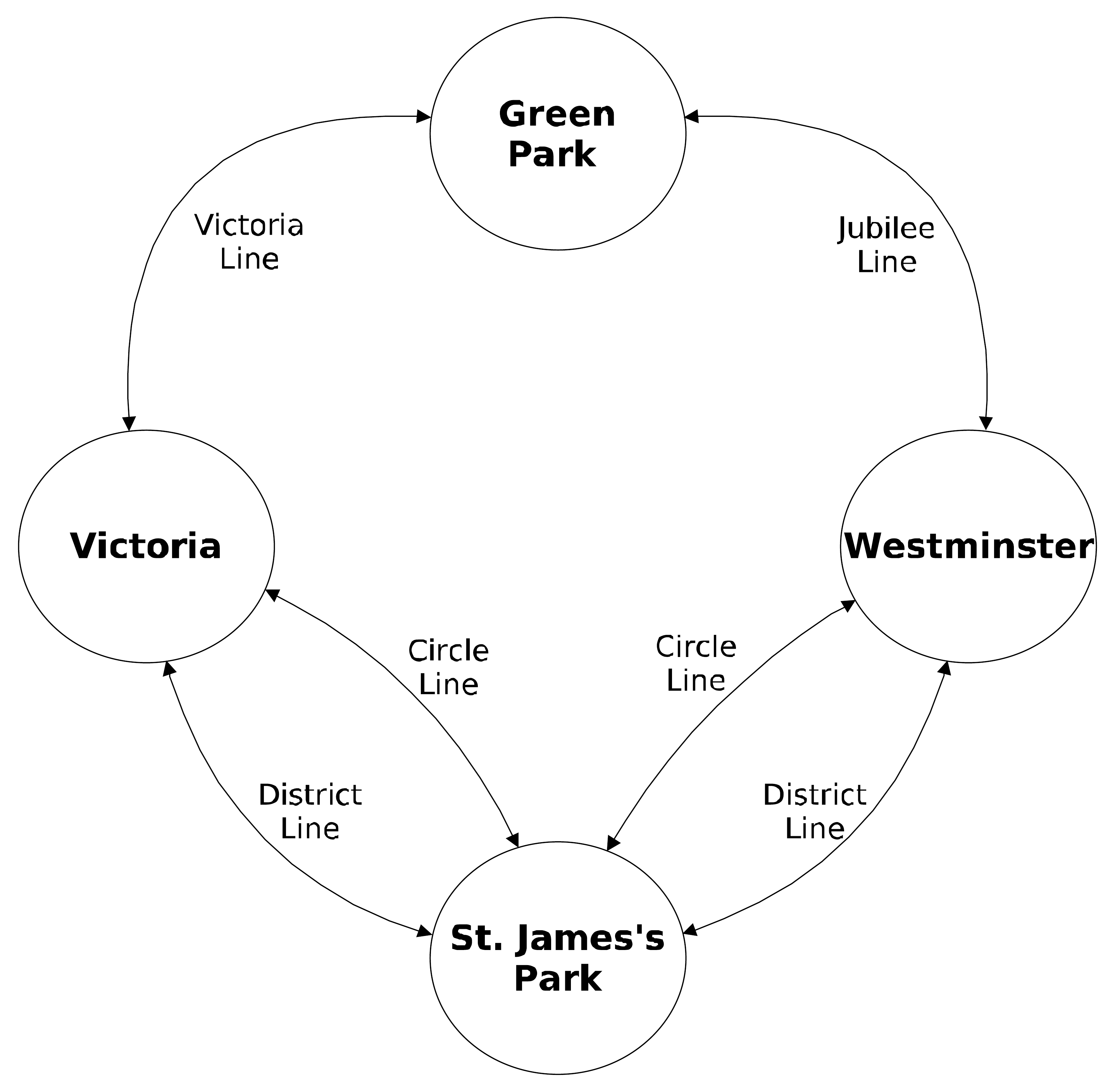}
\centerline{\bf a \hspace{21em} b \vspace*{1ex}} \\ \\
\includegraphics[width=0.4\textwidth]{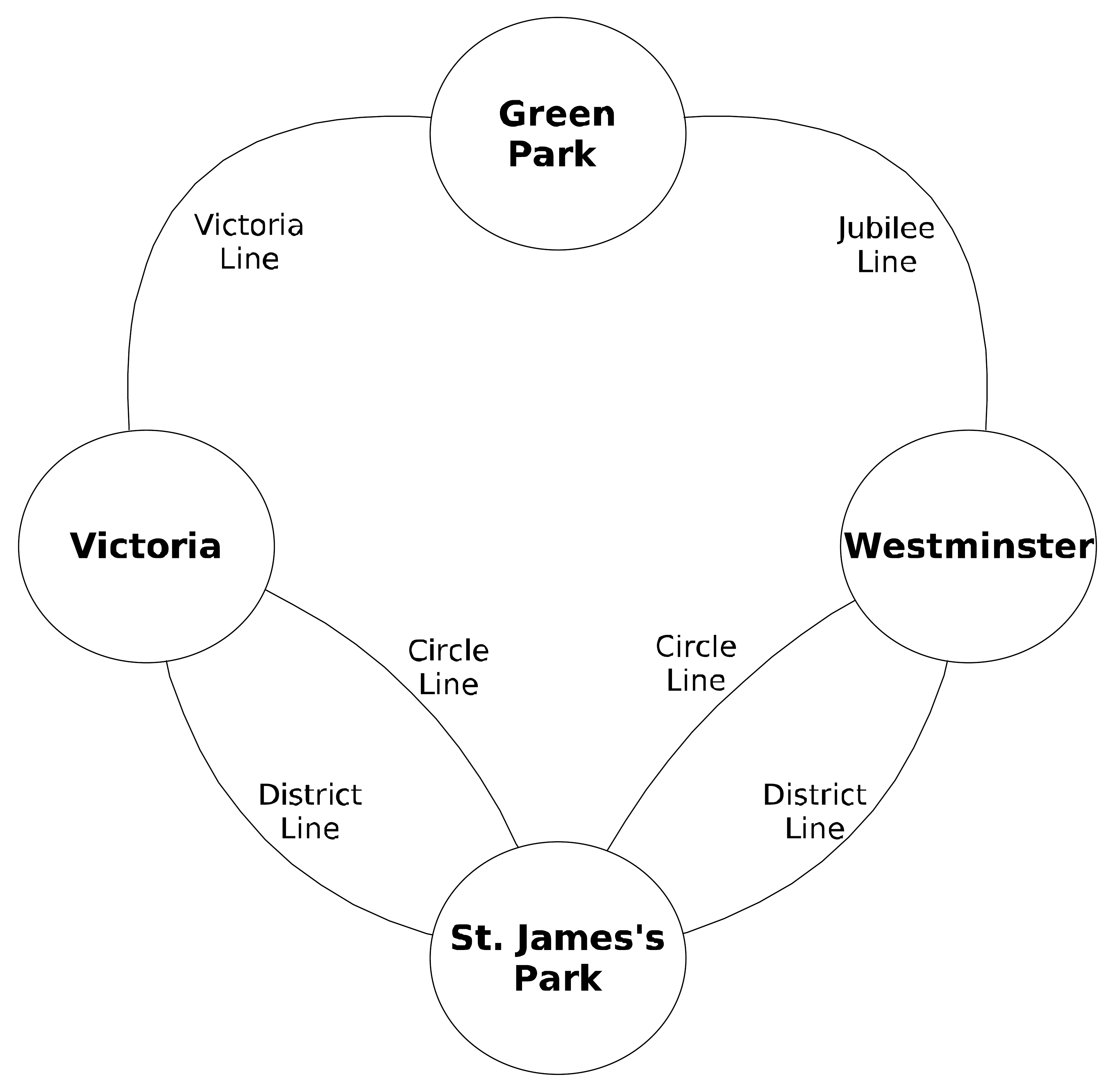}
\includegraphics[width=0.4\textwidth]{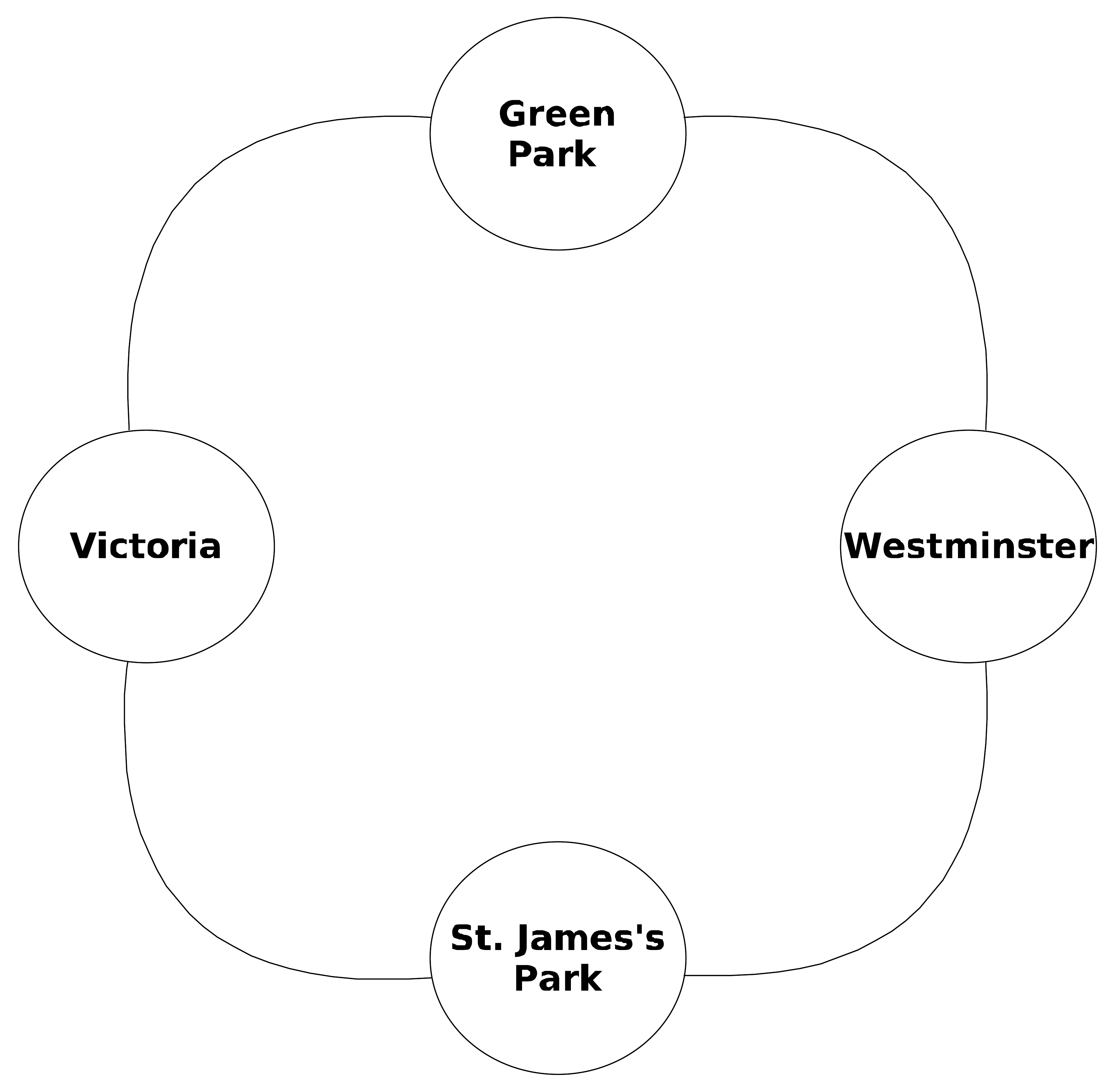}
\centerline{\bf c \hspace{21em} d} \caption{A fragment of the London PTN
and its representations in the form of a graph. {\bf a}: a sample of
the city map that includes several PTN stations (such representation
will be called the geo-space onwards). Lines of different colour on
the map correspond to different PTN routes; {\bf b}: the stations
on map {\bf a} are shown as nodes with links indicating outgoing
and incoming routes; {\bf c}: the same as {\bf b}, but directions
of routes are not shown; {\bf d}: the same as {\bf c} but multiple
links are reduced to single ones. This is the ${\bf L}$-space used
in our analysis.} \label{fig3}
\end{figure}

Although the application of complex network
analysis to the study of PTN has started comparatively recently, sufficient
information has been accumulated to extract some general conclusions. 
Since 2002 when Latora and Marchiori first published their work analysing the topological
properties of the Boston subway (Latora and Marchiori, 2002) many other 
similar studies have been performed all over the world. This can be seen in
Fig \ref{fig1} where black dots indicate approximately where
the topological features of PTN have been analysed. These PTN have 
ranged in size from $152$ to $44629$ stations. The types
of PTN that have been investigated include the subway (Latora and
Marchiori, 2002; Seaton and Hackett, 2004), bus (Xu et al., 2007a; Sui
et al., 2012; Yang et al., 2011; Guo et al., 2013), rail (Sen et al.,
2003), air (Sun et al., 2016, Pien et al., 2015; Guida and Maria, 2007; Guimera et al., 2005; Zang et al., 2010), maritime 
(Xu et al., 2007b; Hu and Zhu, 2009; Lui et al., 2017)  and various combinations of these (von Ferber et al. 2009;
Sienkiewicz and Ho\l{}yst, 2005; Soh et al., 2010;  Zhang et al., 2013, 2014; Alessandretti et
al., 2015; Zhang et al. 2016a, 2016b.\\
Currently, in complex network science a number of topological representations of PTN are in use for the purpose of extracting various types of
information. Different representations may be
implemented by attributing different constituents of the real world
network to graph nodes and edges. For example, one can represent
each PTN station as a graph node and join all
nodes that form part of a particular route to make a complete
subgraph. Different subgraphs will be joined together due to common
stations that are shared by different routes. Such a representation
has been called $\bf P$-space (Sen et al., 2003; Seaton and Hackett,
2004; Sienkiewicz and Ho\l{}yst, 2005; von Ferber et al., 2009; Xu et
al., 2007; Ghosh et al., 2010). It is useful in particular for
determining the mean number of vehicle changes one has to take when
traveling between any two points on the network. In the
so-called $\bf B$-space (von Ferber et al., 2007; Chang et al., 2007)
one constructs a bipartite graph that contains nodes of two types:
node-stations and node-routes. Only nodes of different types can be
linked: a node-station is linked to the node-route if it belongs to
that route. One can pass from such representation to a graph where
only nodes of one type are present. This is achieved by 
a single mode projection, when all nodes of a similar type which are linked
to a common node of another type are represented as a complete subgraph.
Naturally, the single mode projection of the $\bf B$-space graph to the 
nodes-stations leads to $\bf P$-space. In turn, an analogous projection
to the nodes-routes leads to the so-called $\bf C$-space (von Ferber et al., 2009).
Here, one considers how routes are connected to each other. In $\bf C$-space if any
two routes service the same station they are obviously linked.
In Fig \ref{fig3} we show a schematic view of the
situation in the $\bf L$-space (as this is the topology we use 
in this study). 
As it can be seen from Fig \ref{fig3}, the $\bf L$-space
representation is constructed following a simple process. If two
stations are adjacent in a route a link is formed between the two
stations. However, if there are multiple routes going through the
same two stations, $\bf L$-space will not reflect this as it will
not permit multiple links. This topology is ideal for studying the
connectivity of networks for example calculating metrics like mean
path length $\langle \ell \rangle$, Giant Connected Component (GCC)
and other similar metrics, see the Appendix for definitions, here and below.
This space is probably the most commonly used topology and has
been applied in many different studies on PTN
(Latora and Marchiori, 2001;  Sienkiewicz and Ho\l{}yst, 2005; von Ferber et al., 2005; Angeloudis and Fisk, 2006; Xu et al., 2007; Ferber et al., 2009).
These networks can be analysed as a simple graph
with either no weights or as a weighted graph (Latora and Marchiori, 2002).
In Latora and Marchiori (2002) it is argued that weighted networks provide more
realistic information on PTN especially with regard to $\langle
\ell \rangle$. This is because $\langle \ell \rangle$ would
effectively measure the time or distance taken rather than just the
number of stations traveled between two given stations which is
where the unweighted network is losing information. It has been argued in 
Kosmidis et al. (2008) that the spatial embedding (distribution of nodes in Euclidean space) does affect its properties and should be considered when analysing networks that are spatially embedded.
\\
Using networks provides access to different observables quantifying 
general PTN properties: distributions of node degrees, clustering,
assortativity, shortest path length and small-worldedness.
As research has progressed other features have become
of interest for example how different routes tend to show 'harness'
behavior i.e. follow similar paths for a certain number of stations.
This feature was analysed in von Ferber et al. (2005); von Ferber et
al. (2009); Berche et al. (2009), where the harness distribution
$P(r,s)$ defined as the number of sequences of $s$ consecutive
stations that are serviced by $r$ parallel routes. 
A similar feature has been treated for weighted networks in Xu et al.
(2007). Both methods produce power law distributions for their
respective networks.
\\
The criticality of nodes in the international air transportation country networks has been studied in detail (Sun, X., Wandelt, S., \& Coa, X., 2017). Different criticality techniques that account for the unweighted structure of air transportation networks, complex network metrics with passenger traffic as weights, and ticket data-level analysis were applied.
PTN robustness to targeted and random removal of their constituents
(attacks) have also been considered (Sun, X., Wandelt, S., \& Zanin, M., 2017, Pien et al., 2015; Berche et al., 2009; Leu et al., 2010; Berche et al., 2012; von Ferber et al., 2012, Bozza et al., 2017,
Xing et al., 2017). 
One of the goals of these studies is to present criteria, that allow for a priori quantification of the stability of real world correlated networks of finite size and to confirm how these criteria correspond to analytic results available for infinite uncorrelated networks. The analysis focused on the effects that defunct or removed PTN constituents (stations or
joining links) have on the properties of PTN. Simulating different
directed attack strategies, vulnerability criteria have been derived
that result in minimal strategies that have significant impact on these systems.
\\
The above empirical research has revealed that PTN constructed in cities
with different geographical, cultural and historical background
share a number of basic common topological properties: they appear
to be strongly correlated structures with high values of clustering
coefficients and comparatively low mean shortest path values, their
node degree distributions are often found to follow exponential or
power law decay (the last case is known as scale free behaviour
(Barab\'asi and Albert, 1999)). In turn, collected empirical data has
lead to the development of a number of simulated growth models for
PTN. In Berche et al. (2009) interacting self avoiding walks on a
$2$D lattice with preferential attachment rules are applied to
produce similar statistics to real world PTN. In Torres et al. (2011) 
an optimisation model for line planning is discussed considering the 
competing interests in maintaining a quality service whilst minimising costs. 
In Yang et al. (2011) PTN are grown a route for each time step using an ideal $n$-depth
clique topology. In Sui et al. (2012) the optimised growth of a
route is considered by using two competing factors: investors and
clients; clients want the route to be as straight as possible to save
time whereas investors want the routes to meander in order to
collect as many passengers to maximise profits. As it has been shown recently (Louf et al. (2014)), 
the cost-benefit analysis 
accounts for the scaling relations that govern dependency of the PTN
characteristics with the socio-economical features of the undelying region.
\\
In most of the papers cited above the main subject of analysis was
{\em topology} and its impact on the properties of PTN.
This type of analysis has lead to substantial progress in
understanding the collective phenomena taking place on PTN.
For example, the vulnerability of PTN to random failures and
targeted attacks appears to be tightly connected to the distribution
of nodes of high degree (hubs) (Berche et al., 2009; Berche
et al., 2012; von Ferber et al., 2012). Moreover, the analysis of
network topology allows for the singling out of the most important
nodes that control network integrity and to form alternative methods
to construct robust and efficient PTN.

\subsection*{Geospace}

Another essential ingredient to be considered in parallel with
the analysis of PTN topological properties is the spatial embedding
of PTN. There have been far less of these studies when compared to topological studies.
This is mainly due to the lack of available data on the spatial coordinates of PTN. 
The notion of a fractal (non-integer) dimension is often used to quantify the development and growth of cities including their communication and transportation systems. City growth has been shown to exhibit self-similar behaviour, an observation that might imply a
universality of processes that drive city agglomeration and clustering (Batty, 1994; Batty, 2008). Moreover, several physical growth processes that are known to lead to such geometry
(percolation or diffusion limited aggregation) have been exploited to explain such growth in cities (Batty, 1994; Batty, 2008; Makse et al., 1995; Holovatch et al., 2017).
\\
There have been a few studies that directly consider PTN spatial analysis which mainly focus on modes of transport such as railway and subway (Sun et al., 2017; Sui et al., 2012; Benguigui and Daoud, 1991; Benguigui, 1995; von Ferber
and Holovatch, 2013; Guo et al., 2012; von Ferber et al., 2009; Frankhauser, 1990;
Thibault, 1987; Kim et al., 2003) and with the availability of data improving more studies are sure to follow. One of the earliest studies is that of Benguigui and Daoud (1991) measuring the fractal dimension of the Paris subway and railway network by counting the number of stations $N(r)$ within a radius $r$ for a given centre as a function of the radius $r$. In Sui et al. (2012) the end to end mean distance $\langle R \rangle$ 
of routes in nine Chinese cities is computed while in von Ferber et al. (2009) $\langle R \rangle$ is calculated as a function of the number of stations. In von Ferber and
Holovatch (2013) the distribution of these inter station distances are analysed and are found to follow Levy flight distributions. In Sun et al. 2017 they consider how the fractality of worldwide airports network effect network metrics such as nodes, edges, density, assortativity, modularity and communities.
\\
These real world transportation networks have been characterised by varying results (Kim et al., 2003; Benguigui and Daoud, 1991; Benguigui, 1992; Frankhauser, 1990; Thibault, 1987; von Ferber and Holovatch, 2013). In particular, in Thibault (1987) three Lyon regions for rail, bus and drainage networks were shown to have fractal dimensions of ranging between
$1.64-1.88$, $1-1.45$ and $1.21-1.79$ respectively. These fractal dimensions all show that as the radius from the centre of a city 
increases the amount of rail, bus and drainage decreases sub-linearly. Rail displays the largest values of fractal dimensionality 
with the least variance thus indicating that the length of track decreases more slowly than the number of bus stations as the 
distance is increased from the centre of the city. The Stuttgart railway fractal dimension was found to be $1.58$ (Frankhauser, 1990) 
and for the Paris railway the value  $1.47$ was obtained (Benguigui, 1992). The Rhinetowns and Moscow railways exhibited exponents 
of $1.70 \pm 0.05$ and the Paris metro $1.80 \pm 0.05$ (Benguigui and Daoud, 1991). For the Seoul transportation network the exponents 
were measured as $1.5$ for stations and $1.35$ for the railway tracks (Kim et al., 2003).
\\
The majority of the above mentioned papers considered either topological or spatial properties. A particular feature of the study we 
present below is a cumulative analysis of both topological and geographical characteristics. Moreover, to make predictive power of 
our empirical observations more obvious, we complete our study by analysing some features of PTN dynamics and network load. To this 
end we have chosen to consider six GB PTN using the data available on the National Transport Data Repository (Data, 2012). Out of 
those, two PTN operate on an nation-wide scale (national coach and
rail networks) and the remaining four are PTN of Bristol, Manchester, West Midlands and Greater London. By this choice we attempted 
to have examples of areas of different geographical and economical scales. In turn, this enables one to seek  for general (universal) 
characteristics of transportation system as a whole. In the next section we explain the origin of the data and how it will be used in our analysis.

\section{Description of PTN database}\label{II}

\begin{figure}[!h]
\centerline{\fbox{\includegraphics[width=0.90\textwidth]{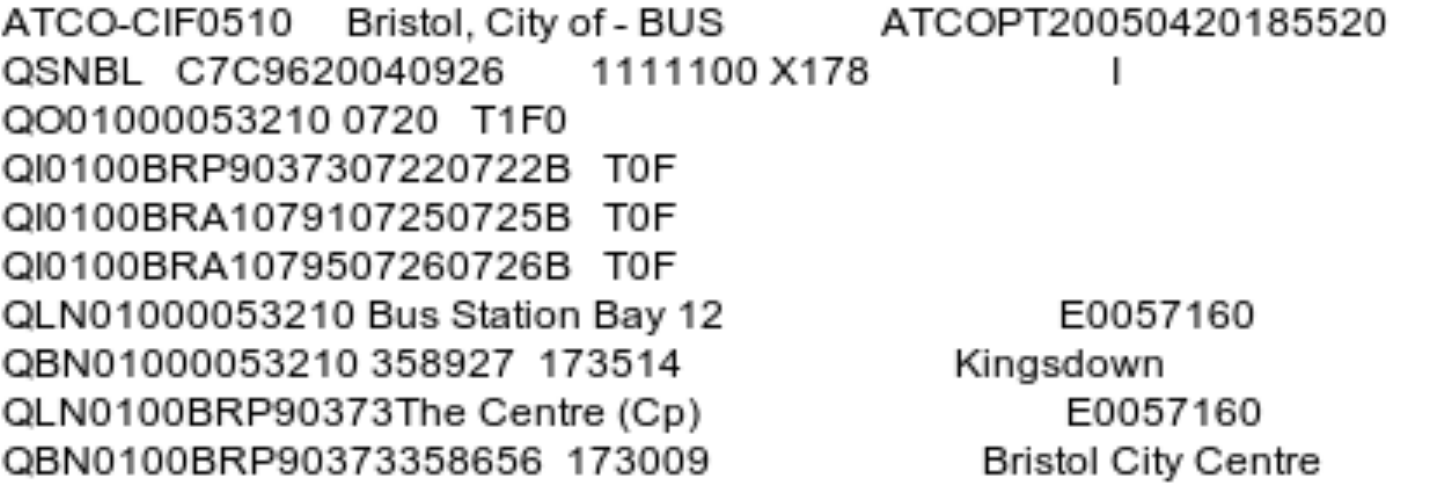}}}
\caption{A snapshot of data
from the Bristol bus network in its raw form. Each line starting
with QS represents a route, each line beginning with QO, QI, QT
represent the start, intermediate and terminal station respectively
for a route. The following 12 characters on these lines are the unique
station identifiers followed by times that the a bus arrive and
leave a station. Correspondence between station identifiers and station name 
can be found using schedules provided in the repository, see also Ref (Data, Bitbucket).  
Lines starting with QL and QB provide information on spatial
coordinates.
\label{fig2}}
\end{figure}

The data for this study originates from the National Transport Data
Repository (NTDR) website (Data, 2012). The website has an Open
Government License meaning it is open to the public and it contains
information on public transport travel and facilities throughout the
GB for the years $2004$ to and including $2011$\footnote{ This data
has not been updated so far. The reference Gallotti and
Barth\'elemy (2015) is based on an older version of 2010.}. The information
provided is a yearly snapshot of the public transport network for a
sample week in each year. The week on which the data is usually
recorded is either the first or second week in October to avoid
recording during school holidays or other seasonal variations which
are at a minimum during this period according to NTDR.
\\
The data is collected and assembled following a decentralised system
where individual regional travel lines (RTL) are responsible for
recording the travel within their allocated districts. These records
are then sent to the NTDR to be collated into one comprehensive
database. There are $11$ RTLs that provide the NTDR with data,
these are: Scotland, North East and Cumbria, North West, Yorkshire,
Wales, West Midlands, East Midlands, East Anglia, South East, South
West and London. The data for national coach and rail are the only data
sets to be compiled centrally. Using a decentralised method for data
retrieval may have benefits especially when it comes to efficiency,
however, it does create more opportunity for errors. For example
duplication of routes and stations on routes that span borders of
two or more RTL. Other complications result from slight differences
occurring in the formatting of the data sent to the NTDR. However, to
prevent such errors the NTDR has an explicit document detailing the
format of the data. Nevertheless, there remain slight differences in
the format which need to be taken into account when analysing the
data. Fig \ref{fig2} is a snapshot of data taken from the Bristol
bus network in its raw form.
\\
The data set includes transport modes for national coach and rail
which span GB mainland. More specifically, it includes bus networks for all
cities in the GB, as well as metro systems for larger Metropolitan areas like 
London, Greater Manchester and the West Midlands. Some of these
networks are subsets of others i.e. one PTN might cover a county and
another a city within that particular county.
For each mode of transport that a city or county offers, which could
be any combination of coach, train, metro, and ferry, a separate
file is held in the records. For each station the type of
information that can be extracted is the following: the location of
the station within a particular route; first, intermediate or last;
number of times a station is visited throughout the day;
geographical coordinates, using an Easting and Northing reference
system; whether the route is incoming or outgoing and which routes
these nodes belong to. There are errors in the data that do require
removing and some missing data that needs to be considered.
However, in general the database provides a rich platform which
we intend to use to analyse the topological and spatial aspects of
PTN in the GB.

\section{Results and Analysis}\label{IV}

 Using the datasets provided by the NTDR the connectivity and spatial features of each PTN were extracted. For purposes of reproducibility 
 the networks generated and including their spatial features have been made public (Data, Bitbucket). Moreover, in Appendix A we have made clear all definitions of metrics used in this analysis. Now with the data at hand we consider the topological, spatial and dynamical features of these PTN.

\subsection*{Topological properties of PTN}

In this study we will be using the $\bf L$-space topology to represent PTN in complex network form, see Fig \ref{fig3}. This most naturally describes the properties of the PTN we are interested in. In this representation, a node in a graph corresponds to a PTN station. Different nodes are linked together when the corresponding stations are subsequently visited by a vehicle.
\\
In the analysis we consider only outgoing routes. The reason for this is that in general the incoming and outgoing stations are usually on opposite sides of the road or
very nearby. So instead of having a directed network one can assume
both incoming and outgoing stations are the same and reduce the
network to an undirected network. This approach allows for a more
intuitive interpretation of the network statistics. For example if
two stations are next to each other but one on the incoming and the
other on the outgoing line then in a directed network they are
actually far apart as the passenger would have to travel all the way
to the beginning of the line and return on corresponding opposite
route to reach the station across the road. This is avoidable in the
case of an undirected network. Using this method would obviously
cause problems if these incoming and outgoing stations were not
close to each other, but we discard such situations as highly
improbable.
\\
Each network can be uniquely described in terms of its adjacency matrix $\hat{A}$
with elements $A_{i,j}=1$ if there is a link between nodes $i$ and
$j$ and $A_{i,j}=0$ otherwise. In turn, based on the adjacency
matrix constructed for each PTN under consideration, we are in the
position to extract the main observables that are commonly used to quantify
network properties. These are summarized in Table \ref{tab1}
\begin{sidewaystable*}
 {
 \caption{General statistics for PTN under consideration and main topological features of their representations in a form of complex networks. $n$, $m$: number of nodes and links of a PTN network with $R$ routes.
 $\langle k \rangle$, $\langle k^2 \rangle$, $k_{\rm max}$: mean,
 mean square and maximal node degree. GCC: size of the giant
 connected component in percents to the general network size.
 $\langle \ell \rangle$, $\langle \ell_{\rm rand} \rangle$: mean
 shortest path length of a PTN and of a random graph of the same
 size, $\ell_{\eta}$: Measure of efficiency in terms of path length, $D$: diameter (maximal shortest path length). $\langle \ell_t
 \rangle$: mean shortest travel time, in minutes. $r$: assortativity.
 $C$, $C_{\rm rand}$: mean clustering coefficient of a PTN and of a
 random graph of the same size. 
  \label{tab1}}
 \begin{tabular}{lccccccccccccccc}
 \hline \hline
 Networks & $n$ & $m$ & $R$ & $\langle k \rangle$ & $\langle k^2 \rangle$ & $k_{\rm max}$ & GCC  & $\langle \ell \rangle$ & $\langle \ell_{\rm rand} \rangle$ & $\ell_{\eta}$ & $D$ & $\langle \ell_t \rangle$  & $r$  & $C$  & $C_{\rm rand}$  \\ \hline
 Coach  & $2499$ & $3228$ & $258$ &    2.58  & 12.17 & $43$ & $96.1$  & $23.1$ & $7.22$ & $3.2$ & $121$& $359.2$ & $0.27$ & $0.075$ & $1\times10^{-3}$   \\
 Rail  & $2575$ & $4450$ & $504$ &     3.46  & 20.03 & $31$ & $99.9$   & $11.82$ & $5.76$ & $2.1$ & $48$ & $162.9$ & $0.24$ & $0.309$ & $1\times10^{-3}$   \\
 Bristol & $2580$ & $3101$ & $172$ &   2.40  & 6.58 & $10$ & $99.8$ & $27.7$ & $7.92$  & $3.5$ & $122$ & $40.0$  & $0.26$ & $0.034$ & $9\times10^{-4}$  \\
Manchester & $10742$ & $12275$ & $862$ &     2.29  & 5.87 & $21$ & $100$    & $48.7$ & $10.10$ & $4.8$ & $238$& $61.0$  & $0.20$ & $0.028$ & $2\times10^{-4}$  \\
 West Mid & $11743$ & $15034$ & $521$& 2.56  & 7.75 & $20$&$100$ & $52.8$ & $9.06$  & $5.8$ &$168$& $55.4$  & $0.13$ & $0.035$ & $2\times10^{-4}$  \\
 London & $16397$ & $18496$ & $767$ &  2.26  & 5.56 & $11$ &$99.8$ & $53.5$ & $10.82$ & $4.9$ & $206$& $72.5$  & $0.21$ & $0.014$ & $1\times10^{-4}$  \\ \hline \hline
 \end{tabular}}
 
 \end{sidewaystable*}
\\
The first two columns of the table give the number of nodes $n$ and links $m$ for each network,
where the number of nodes directly corresponds to the number of PTN stations. The number of links in ${\bf L}$-space gives a reduced value
of real linkage between the stations, cf. Fig \ref{fig3}. In the table we also display the
number of routes $R$ for each PTN, this does not have its counterpart in network
topology for ${\bf L}$-space. The number of links adjacent to a given node $i$ is
called the node degree, $k_i$. It serves as one of the indicators to show the importance of
a node in the network. Defined in terms of the adjacency matrix it reads:
\begin{equation}\label{3.1}
k_i=\sum_{j} A_{ij}\, ,
\end{equation}
where the sum is taken over all network nodes. Table \ref{tab1}
gives mean $\langle k \rangle$, mean square $\langle k^2 \rangle$ and maximal values $k_{\rm max}$ of node degrees for each of the networks. One important 
anomaly to mention with the data is the London PTN results differ slightly from those presented in von Ferber et al. (2009). The main reasons for the 
discrepancy is that we only consider one mode of transport (bus) whereas von Ferber et al. studied multiple modes of transport. Further to this bus networks
have the ability to evolve quicker than other forms of transport due to the relatively low cost involved. Paper by von Ferber et al. 2009
considered the network for the year 2007 whereas we studied the network for the year 2011. Over this time frame some changes may have occurred 
resulting in the discrepancy we see in the results between the two networks.
\\
\subsection*{Topological measures of robustness}

Obviously, network integrity plays a crucial role in various processes occurring on the
network. In particular, transportation can not be maintained between nodes belonging to different
network fragments that are not joined together. As one can see from the table, the largest connected component
of each PTN (giant connected component, GCC) includes almost all nodes, making any location
on the network reachable from any other location. Slight deviations below 100 \% are caused either by quality of data or by fact that some PTN are operated by different companies not represented in the database. Moreover stations are assigned according to geographical location and for example on a busy road servicing many routes, stations could be placed next to each other and not be linked, generating a segmentation in the network.
\begin{table}[th]
\begin{center}
\tabcolsep1.2mm
 {\small
 \caption{Molloy-Reed parameter $\kappa$ (\ref{3.2}) for several PTN
 of the GB cities (our data: first three lines of the table) in
 comparison with PTN of some other cities of the world, as obtained
 in (Berche et al. 2009). The type of transport taken into account are:
 Bus:B; Electric trolleys:E; Ferry:F; Subway:S; Tram:T; and Urban train:U. 
 PTN size is given in terms of number of
 stations $n$ and of routes $R$. The next column gives an exponent $\gamma$
 in the power law (\ref{3.6}) fit, bracketed values indicate less
 reliable fits, see text. The last column is the ratio, $R_{\rho}$ between $n$ and $R$. \label{tab2}}
\begin{tabular}{lcrrrrr}
\hline\hline
City           & Type  &  $n$  & $R$   &$\kappa$& $\gamma$ & $R_{\rho}$\\
\hline
Dallas         & B     & 5366  &   117 &   2.35 &  5.49 &  45.86  \\
London         & B     & 16397 &   767 &   2.46 &  4.25 &  21.38  \\
West Midlands  & B     & 11743 &   521 &   2.56 &  3.10 &  22.54  \\
Manchester     & B     & 10742 &   862 &   2.56 &  4.36 &  12.46  \\
Istanbul       & BST   & 4043  &   414 &   2.69 &  4.04 &  9.72  \\
Los Angeles    & B     & 44629 &  1881 &   2.73 & 4.85  &  23.73 \\
Bristol        & B     & 2580  &   172 &   2.74 &  3.56 &  15.00 \\
Berlin         & BSTU  & 2992  &   211 &   3.16 & (4.30)&  14.18 \\
D\"usseldorf   & BST   & 1494  &   124 &   3.16 & 3.76  &  12.00 \\
Hamburg        & BFSTU & 8084  &   708 &   3.26 & (4.74)&  11.33 \\
Rome           & BT    & 3961  &   681 &   3.67 & (3.95)&  5.87 \\
Taipei         & B     & 5311  &   389 &   4.02 & (3.74)&  13.6 \\
Sydney         & B     & 1978  &   596 &   4.37 & (4.03)&  3.31 \\
Hong Kong      & B     & 2024  &   321 &   5.34 & (2.99)&  6.31 \\
Sa\~o Paolo    & B     & 7215  &   997 &   5.95 & 2.72  &  7.23 \\
Paris          & BS    & 3728  &   251 &   6.93 & 2.62  &  14.84 \\
Moscow         & BEST  & 3569  &   679 &   7.91 & (3.22)&  5.23 \\
\hline \hline
\end{tabular}
}
\end{center}

\end{table}
\begin{figure}
\includegraphics[scale=0.4]{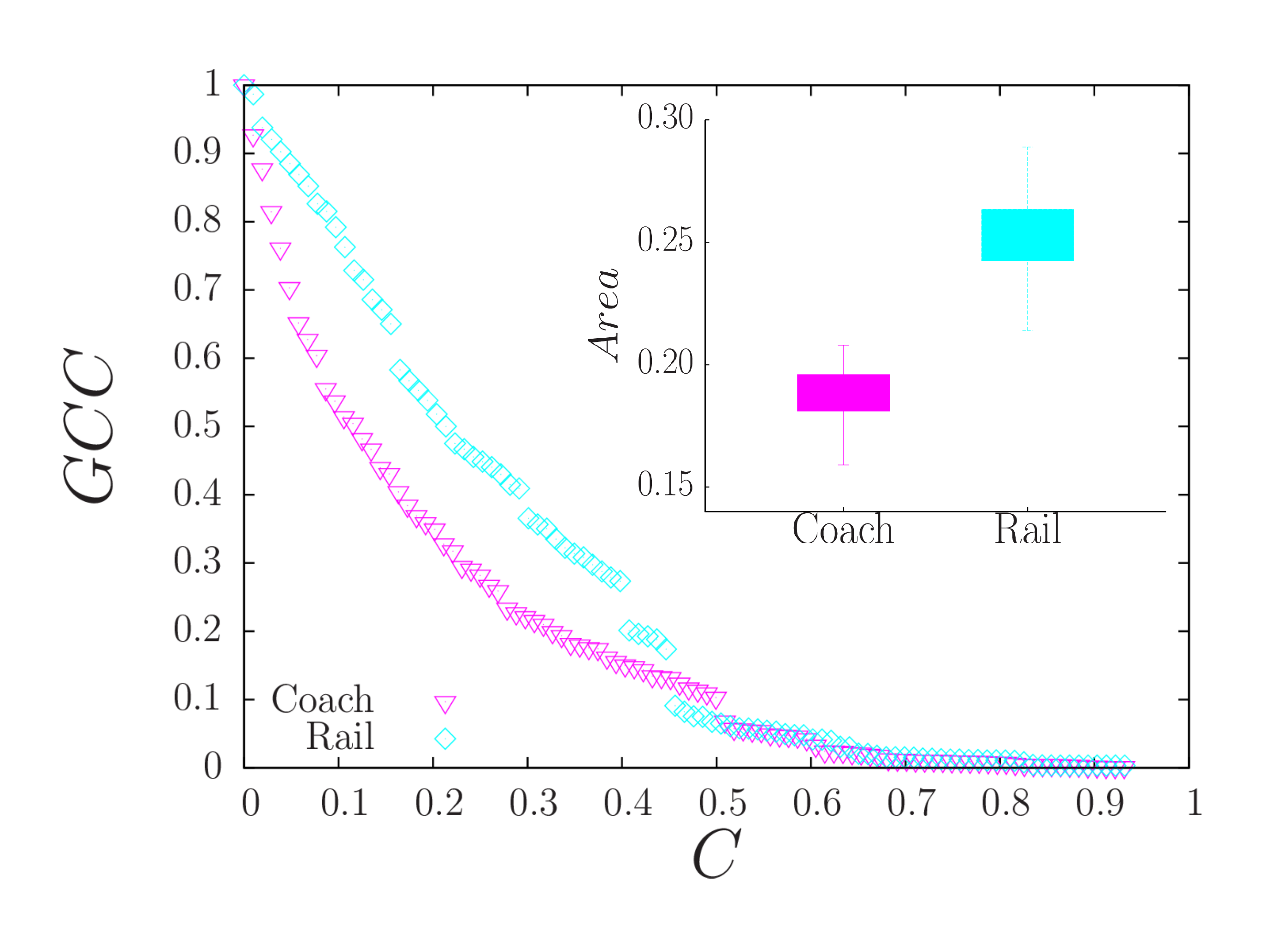}
\caption{The normalized size of the largest connected component GCC of the coach and rail PTN as function of the share $c$ of removed randomly chosen stations. The insert represents the distribution of network robustness, measured as the area under the curve, for 100 simulations of random failure.}\label{Random_Resilience}
\end{figure}
\\
The analysis of topological features of real-world networks can be used
to predict their behaviour under removal of their constituents.
Such removal, usually named an attack or failure, may address network
nodes or links and may be performed at random (random failure) or
may be targeted at the most important components in the network (targeted
attack).
\\
A useful criterion in determining the network
vulnerability is known as the Molloy-Reed criterion (Molloy and Reed
1995). It states that in any uncorrelated network the GCC is present
if:
\begin{equation}\label{3.2}
\kappa= \langle k^2 \rangle  /\langle k \rangle  \geq 2 \, .
\end{equation}
The Molloy-Reed parameter $\kappa$ allows for the evaluation of network
stability to random failures. The higher the value of $\kappa$, the
more stable the network, i.e. the higher the number of nodes that
should be removed to destroy a given GCC. Although  Eq. (\ref{3.2}) has been
obtained for infinite uncorrelated networks as we will see below, it provides useful information on the robustness of real world PTN of finite size. To this end,
in table \ref{tab2} we compare values of $\kappa$ for several GB cities obtained by us
and for PTN in other cities around the world.
The table is ordered in ascending order $\kappa$, column $5$. The higher the value of $\kappa$, the
more stable the network, i.e. the higher the number of nodes that
should be removed to destroy a given GCC. Although  Eq. (\ref{3.2}) has been
obtained for infinite uncorrelated networks, it provides useful information on the 
robustness of real world PTN of finite size, Berche et al., 2012. To this end,
in table \ref{tab2} we compare values of $\kappa$ for several GB cities obtained by us
and for PTN in other cities around the world.
The table is ordered in ascending order $\kappa$, column $5$. The higher the value of $\kappa$, the more robust the PTN with respect to random 
removal of its nodes. From the table we can see that all GB PTNs analysed here feature in the top seven cities with the smallest value of $\kappa$ tending towards $\kappa=2$. As we noticed, for an infinite uncorrelated network such a result indicates that the network is close to its 'percolation' limit, i.e. when the GCC ceases to exist. For a real world network however, such a result serves rather as evidence of an impact on the network topology as its stability tends to be lower. This is caused both by the fact that the real world networks considered here are of finite size and the correlations that are obviously present in their structure. The last assumption is further supported by the values of PTN clustering coefficients, see table \ref{tab1}. A possible explanation of such an effect is found when comparing values of the node degree distribution exponent $\gamma$ (see Eq. (4) for the definition) for different PTN. A high value of $\gamma$ found for some networks considered here brings about a low impact on highly connected nodes (hubs) for which the latter are important for keeping a network connected.
Moreover, observing table \ref{tab2} we can see that the relative size of network routes 
may be also a contributing factor to its stability. This seems to indicate that while these networks are similar in structure investigating more subtle features can highlight important contributing factors to their stability, see also Neal (2004). Indeed, most of the PTN with the lowest $\kappa$ value are characterised by the larger value of a mean route size $\rho$. \\
In order to demonstrate the reaction of PTN with different values of $\kappa$ to random attacks, it is instructive to observe robustness of national networks that cover larger but similar geographic areas then local networks. Here we get values of $\kappa=4.72$, and $\kappa=5.79$ for national coach and rail networks respectively.
According to the Molloy-Reed parameter the national rail has to be the more stable of the two PTN. To prove that this simple and easily evaluated parameter does indeed provide accurate measures of robustness we have simulated $100$ random failures on both national PTN and determined their average robustness. This was performed by determining the area under the curve generated by random failure as we can see in Fig $4$. There, we plot the normalised size of the largest connected component of national PTN as function of the share of removed, randomly chosen stations. Qualitatively we can see that the rail network is more resilient than coach and in the insert this is further confirmed by observing that the robustness distribution is qualitatively different with national rail being more robust then its national counterpart. 
More generally this then supports the idea that indeed for a real-world network an increase in the value of the Molloy-Reed parameter corresponds 
to a more stable network. It will be interesting to check these values against their counterparts for the
networks covering larger geographic space in other regions of the world.
\\
One of the indicators to measure the distance between nodes, providing a useful measure of the efficiency of a PTN, is given by the mean shortest path length $\langle \ell
\rangle$. It is measured by the smallest number of steps one has to
traverse from one node to another given node.
It is instructive to compare properties of the networks under consideration with those of the Erd\"os-R\'enyi classical random graph of the same size, i.e. when the same number of nodes $n$ are randomly linked together by $m$ links. To do this we simple calculate $\ell_{\eta} = \langle \ell \rangle / \langle \ell_r \rangle$. It can be seen in table \ref{tab1} that larger PTN tend to be less efficient than their smaller counterparts.
\\
Specific forms of correlation which are often present in real world
complex networks are measured by the clustering coefficient $C$. It
reflects how many nearest neighbours of a given node are nearest
neighbours of each other. To give examples, $C=0$ for a
tree-like network and $C=1$ for a complete graph, when all nodes are
interconnected by direct links. Usually, $d$-dimensional regular
structures possess high correlations, whereas random structures like the
Erd\"os-R\'enyi graph are characterised by very low values of $C$.
The comparison of data for PTN clustering coefficients $C$ with that of
the classical random graph of the same size $C_{\rm rand}$ gives
undoubted evidence of strong correlations in PTN: $C/C_{\rm
rand} \sim 10^2$ almost for all networks.
\\
Many of natural and man-made complex networks are the so-called ``small worlds''. Being highly correlated, they are characterized by small typical distance, as random structures. 
When considering small worldedness as defined by Watts and Strogatz (1998):
$C \gg C_{rand}$ and $\langle \ell \rangle \approx \log n$, where $n$ is the number of nodes.
One can see from table \ref{tab1}, the first condition for strongly correlated networks definitely holds.
However, the networks exhibit comparatively large mean shortest path lengths when comparing random networks of a similar size: $\langle \ell \rangle  > \langle \ell_{\rm rand} \rangle$.
Therefore, caution is to be taken when attributing small world
properties to PTN. This may be understandable as many nodes of
degree two exist in PTN.
\\
Another useful observation is that PTN of Manchester, West Midlands and London
all have fairly similar values of $\langle \ell \rangle$: $48.7$,$52.8$ and
$53.5$ respectively, even though London is a much larger city and has
far more stations than the other two networks thus indicating
that the London PTN is more efficient in terms of topology 
than the other two PTN. This may however also reflect that there are competing
interests between network stability and efficiency considering London has the lowest
$\kappa$ value of GB networks.
\\
It is instructive also to calculate the mean shortest path for the 
weighted PTN, attributing to each network link a weight
indicating the time necessary to spend traveling along this link. In
this case, such a mean shortest path for weighted networks, $\langle
\ell_t \rangle$ (where edge weights are derived from the NTDR
as described in the caption to Fig. 3) indicates the mean time needed to traverse the
network. As one can see from table $1$, for national networks on
average it takes more than twice as much time to get to any other
station within the network on coach as it does on rail.
\\
Correlation between degrees of neighbouring nodes in a network are
usually measured in terms of the mean Pearson correlation
coefficient $r$. Networks where degrees of the same order tend to be linked together are called assortative, for them $r>0$  and dissortative ($r<0$) otherwise. The
values of $r$ found in our study although being small clearly are in
favour of assortative mixing: as one can see from table $1$,
$r=0.1-0.3$ for the networks under consideration. This means that
edges tend to connect nodes of similar degree. This is not always 
the case for PTN as it has been found
in von Ferber et al. 2009, that some large cities (as
D\"usseldorf, Moscow, Paris, Sa\~{o} Paolo) show no preference in
linkage between nodes with respect to node degrees ($r\simeq 0$). So
in this respect PTN analysed in our study belong to the group of
that includes Berlin, Los Angeles, Rome, Sydney, Taipei
($r=0.1-0.3$) (von Ferber et al. 2009).
\\
It is worth noting another observation that follows from the table
\ref{tab1}: although it includes PTN that span over quite different
distances in the geographic space, their topological features
manifest striking similarities! Indeed, all the networks considered
in this study possess comparatively low value of the mean node
degree, high clustering coefficient, they are disassortative with
respect to node-node correlations. Moreover, the presence of high
clustering in these networks is not accompanied by a low value of
the mean shortest path length, as it usually is expected for the
small world networks.
\\
All together, the above calculated observables characterise the topological features of each of the PTN in a unique and comprehensive way. In turn, this enables comparison of the networks under consideration with other PTN on a base of solid quantitative criteria. Such observables can be employed as key performance indicators (KPIs) in aid of further developing efficient and stable PTN.

\subsection*{Degree distribution}

The node degree distribution $P(k)$ gives the probability to find in a
network a node of given degree $k$. Very often for complex networks its decay is governed by exponential or power laws: 

\begin{equation}\label{3.5}
P(k) \sim e^{-\xi k}\, ,
\end{equation}

\begin{equation}\label{3.6}
P(k) \sim k^{-\gamma} \, ,
\end{equation}

at $k \gg 1$. Here $\xi$ and $\gamma$ are the exponents that describe an exponential and
power law decay respectively.

 \begin{figure}
 \centerline{
   \includegraphics[width=0.5\textwidth]{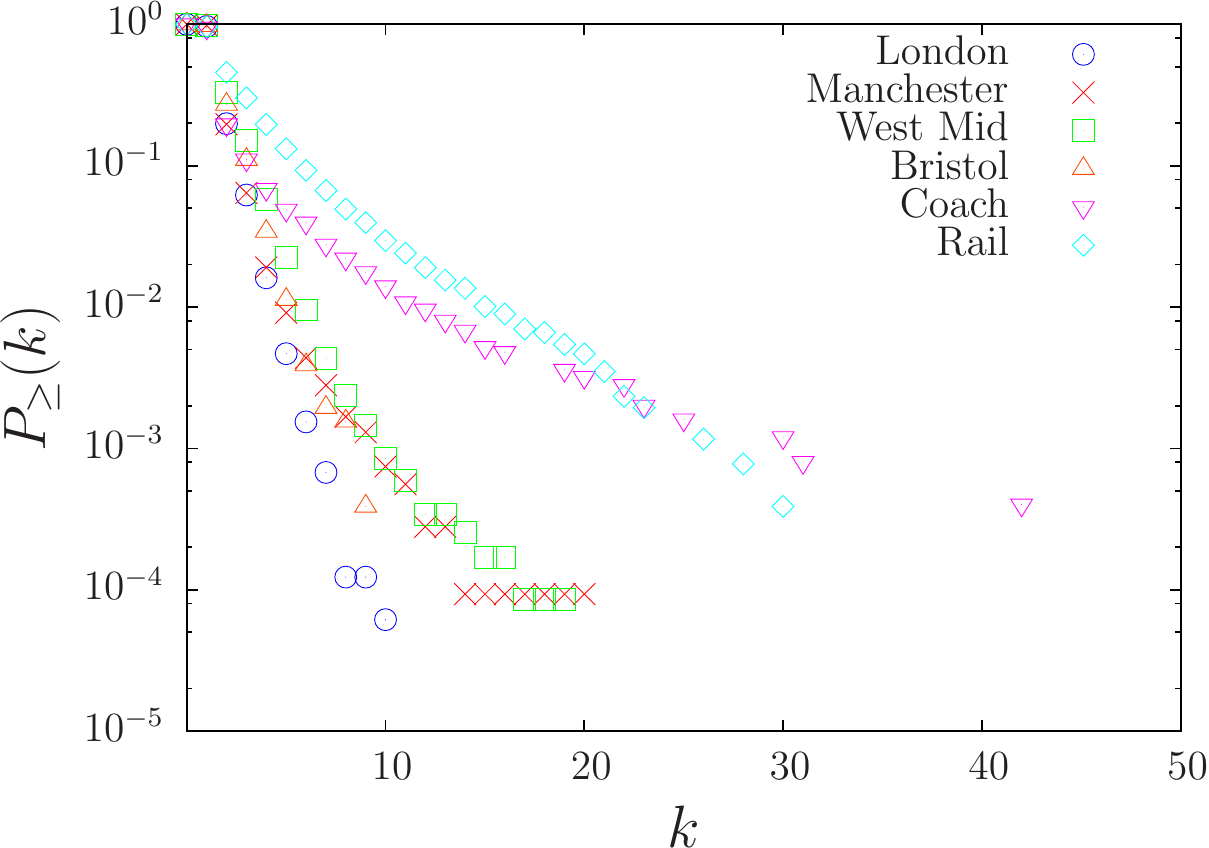}
   \includegraphics[width=0.5\textwidth]{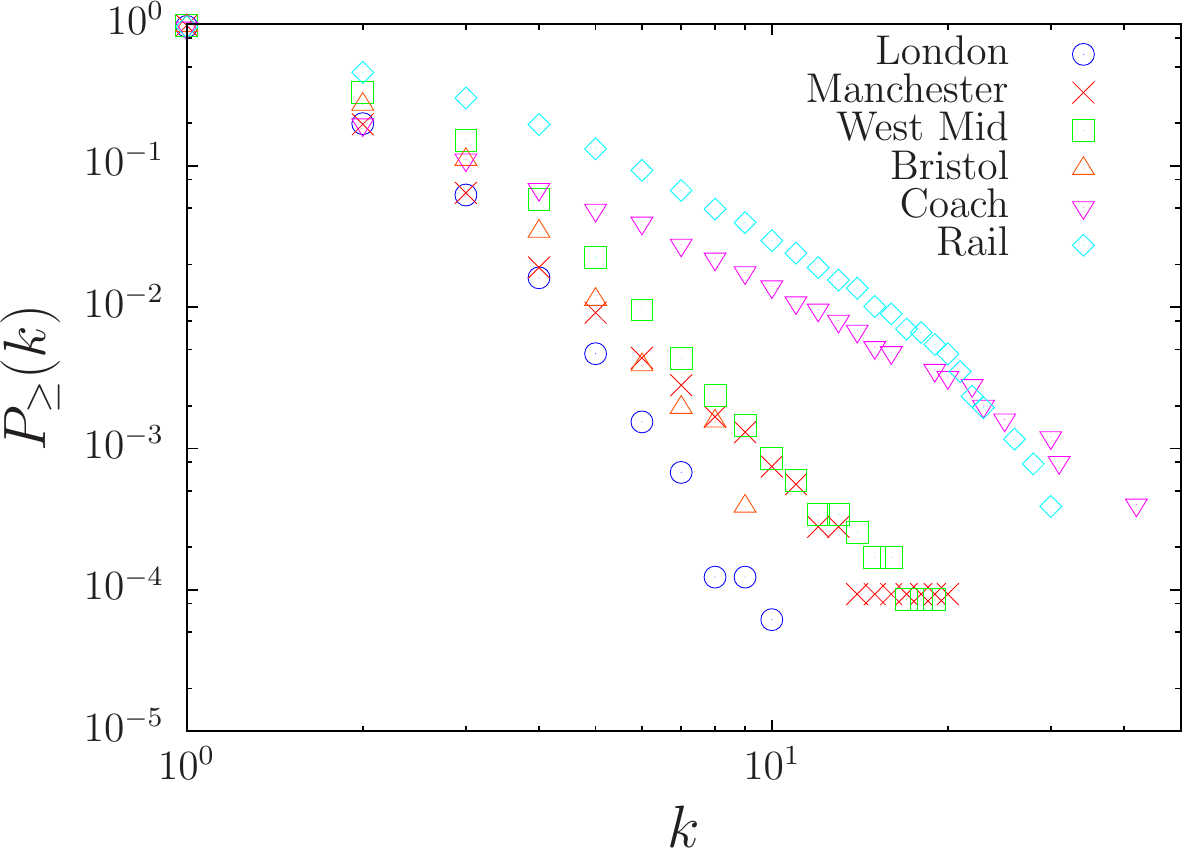}}
   \caption{Cumulative degree distribution $P_\geq(k)$ for six PTN under consideration  in log-lin (left) and log-log (right) scales.}
   \label{fig4}
 \end{figure}

In order to gain access to the $P(k)$ dependencies
Eq.(\ref{3.5}), Eq.(\ref{3.6}) we first plot in Fig \ref{fig4}  corresponding curves
for the cumulative distributions:
\begin{equation}\label{9}
 P_{\geq}(k) = \sum_{q=k}^{k_{\rm max}} P(q)
 \end{equation}
where $k_{\rm max}$ is the maximal node degree for the given PTN.
The cumulative distributions are generally known to behave smoother
and their functional dependence enables a more accurate
determination of  $P(k)$. Corresponding cumulative distributions are
shown in Fig \ref{fig4} both in the log-linear and in the double
logarithmic scales. The exponential dependency (\ref{3.5}) will be
reflected as a straight line in the log-linear scale, whereas the
power law (\ref{3.6}) corresponds to the straight line in the double
logarithmic scale. On inspection it seems that the degree
distributions of these networks show clear preference with respect
to the power law decay. For confirmation, using a nonlinear
least-squares (NLLS) Marquardt-Levenberg algorithm (Fronczak and
Ho\l{}yst 2004; Levenberg 1944), we have produced the fits for these
distributions and display the fitted values of $\xi$ and $\gamma$ in
Table \ref{tab3}. To determine the best fitted function we integrate the upper and lower bounds for the error bars in linear space and subtract these areas. The function that produces the smallest area after this operation is deemed the best fit. For a more detailed explanation of this method the reader is referred to Appendix B. As it follows from our analysis, the node degree
distributions are better fitted by the power-law (\ref{3.6}) than by
the exponential decay (\ref{3.5}).

 \begin{table}[!htb]
{
 \centering
  \caption{Fitted degree distribution exponents $\xi$ (\ref{3.5}) and $\gamma$ (\ref{3.6}).
 For all PTN considered here, the $P(k)$ dependency is better
 fitted by the power-law (\ref{3.6}) than by the exponential decay
 (\ref{3.5}). \label{tab3}}
  \begin{tabular}{ccccccc}
  \hline \hline
 & Coach  & Rail & Bristol & Manchester  & West Mid  & London  \\
  $\xi$ & $2.08 \pm 0.09$ & $0.90 \pm 0.06$ & $1.37 \pm 0.07$ & $1.73 \pm 0.05$  & $1.15 \pm 0.05$& $1.67 \pm 0.05$  \\
  $\gamma$ & $5.06 \pm 0.15$  & $2.5 \pm 0.07$ & $3.56 \pm 0.09$ & $4.36 \pm 0.04$ & $3.10 \pm0.07$  & $4.25 \pm 0.05$  \\ \hline \hline
   \end{tabular}}
 
  \end{table}

Complex networks with clear power law decay of the node degree
distribution are named scale-free. Although we can not attribute clear
scale free features to all PTN examined in this study, the data
displayed in Table \ref{tab3} reports of a power law decay
tendency of the networks under consideration. When this is the case,
the networks with a lower value exponent $\gamma$ should
manifest stronger stability with respect to the removal of their
constituents (see also table \ref{tab2}). A prominent example
follows from the comparison of the GB national coach and rail
networks: the $\gamma$ exponent for the rail PTN is only half that
of than  its coach counterpart. This brings about a higher
stability of the former under random removal of its constituents.

\subsection*{Geospatial properties of PTN}
\begin{figure}[!htb]
\centerline{
\includegraphics[width=0.8\textwidth]{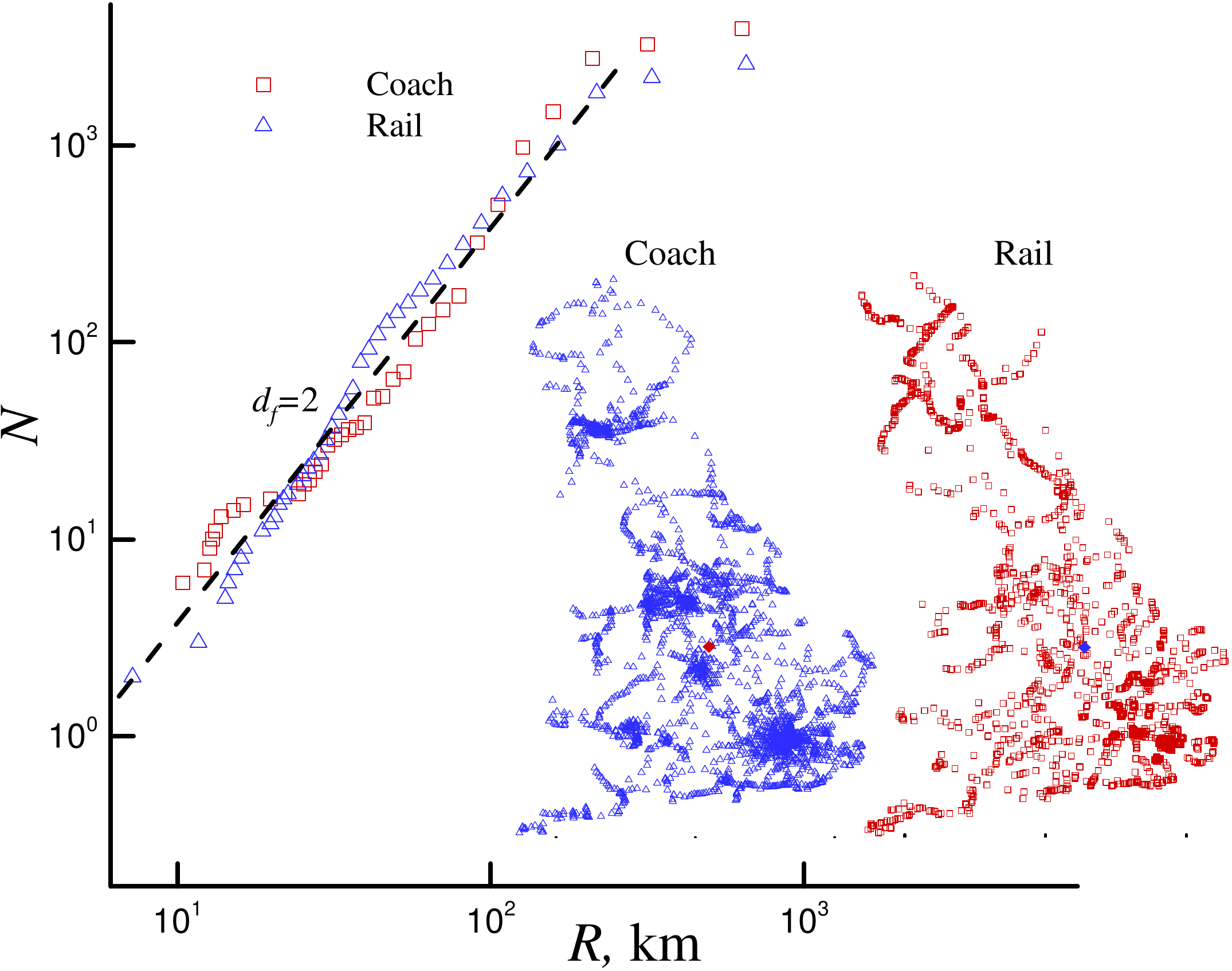}}
\caption{Number of stations $N(R)$ in the circle of radius $R$ for
the GB national coach and rail networks. The fit of the dependence
to the straight line on the log-log plot brings about the exponent
close to $d_f=2$ (dashed line). The inset shows the networks in
geospace.
 \label{fig6}}
\end{figure}
\begin{figure}[!htb]
\centerline{
\includegraphics[width=0.9\textwidth]{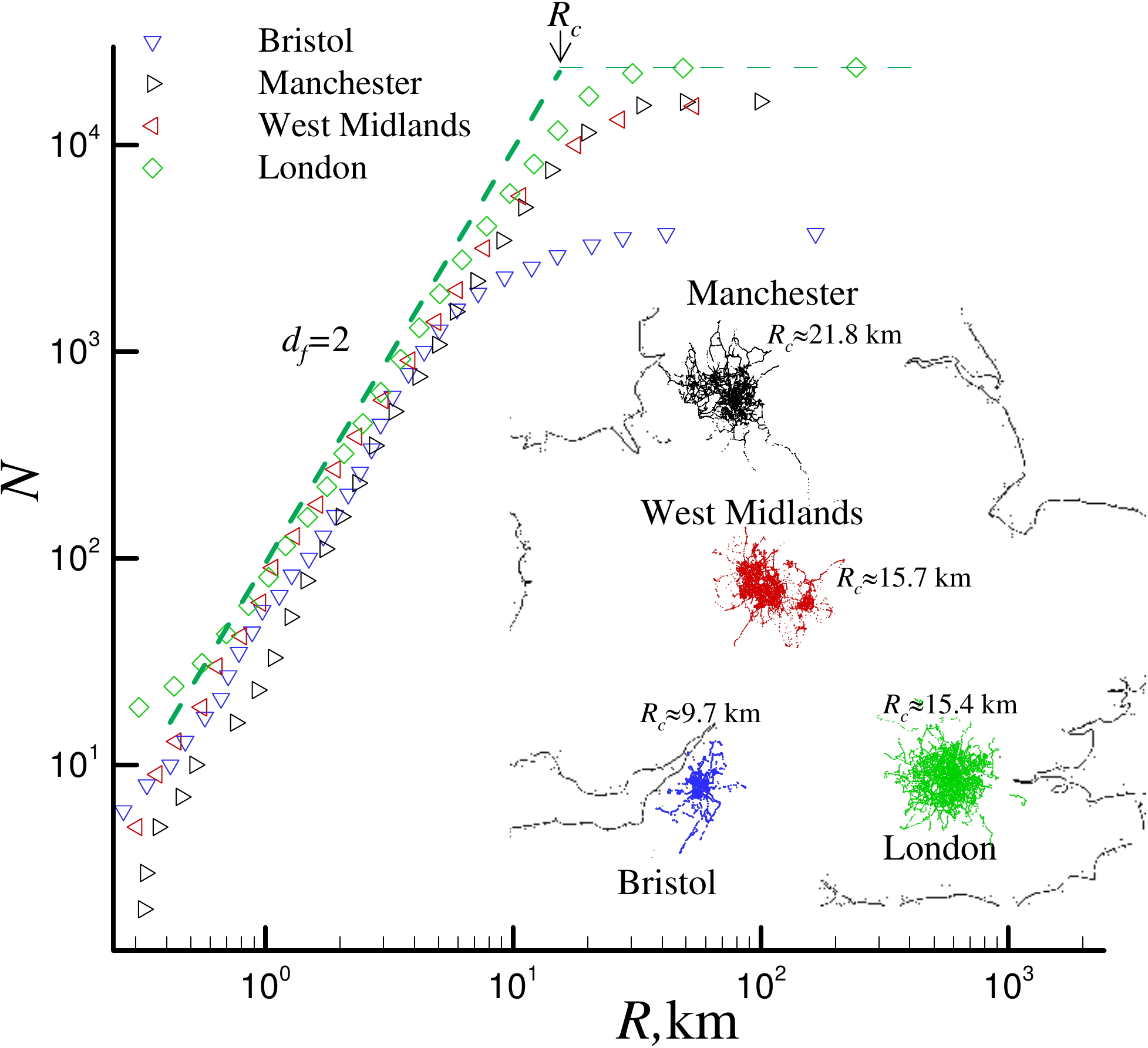}}
\caption{Number of stations $N(R)$ in the circle of radius $R$ for the 
PTN of Bristol, Greater Manchester, West Midlands, and Greater
London. The bold dashed line (green online) corresponds to the
exponent $d_f=2$. The inset shows the networks in geospace, together
with a part of the GB coastline. The radius $R_c$ corresponds to the
transition from the compact central area to the rarefied space with
$d_f<2$. The value of $R_c$ is shown for each network. \label{fig7}}
\end{figure}

Thus far we have been investigating PTN properties that originate from their
topology. Very often data on network topology is not accompanied by their
location in embedded Euclidean space. The advantage of the
database we are using is that it contains the geographical coordinates of stations.
This gives us the unique possibility to complement the topological analysis by examining properties in the Euclidean two-dimensional ($d=2$) space, for which we will call geospace from here onwards. This neglects the slight curvature in the earth but does not effect calculations over the area considered in our analysis. In this section we will be
interested in the spatial distributions of nodes. Insets in Figs \ref{fig6}, \ref{fig7} display the positions of PTN in geospace. It is the distribution of these co-ordinates that will be of interest in this section.
\\
Analysing the fractal dimension of PTN two methods have been
considered in this study each providing different but useful interpretations on serviceability of PTN. In turn, this opens up a method to use fractal dimensionality as a KPI, giving one more quantitative characteristic of a PTN functional effectiveness.
Initially we find the centre of mass and
investigate the "mass" (number of stations) of the network $N(R)$ as a function of the
radius $R$ about this centre. This is done within the distance range 100~m~$-$~100~km for local networks and 1~km~$-$~600~km for national networks. If the scaling (power law dependence)
\begin{equation}\label{4.1}
N(R)\sim R^{d_f} \ ,
\end{equation}
is observed with a non integer value of the exponent $d_f$, the
exponent is associated with the fractal dimension of the network.
Indeed, if the stations in the PTN were equidistantly distributed
along straight lines in a one dimensional case, this would correspond
to the exponent $d_f=1$. Likewise, for a two dimensional case, constant
station density (number of stations per unit area) would lead to $d_f=2$.
\begin{figure}[!htb]
\centerline{
\includegraphics[width=0.9\textwidth]{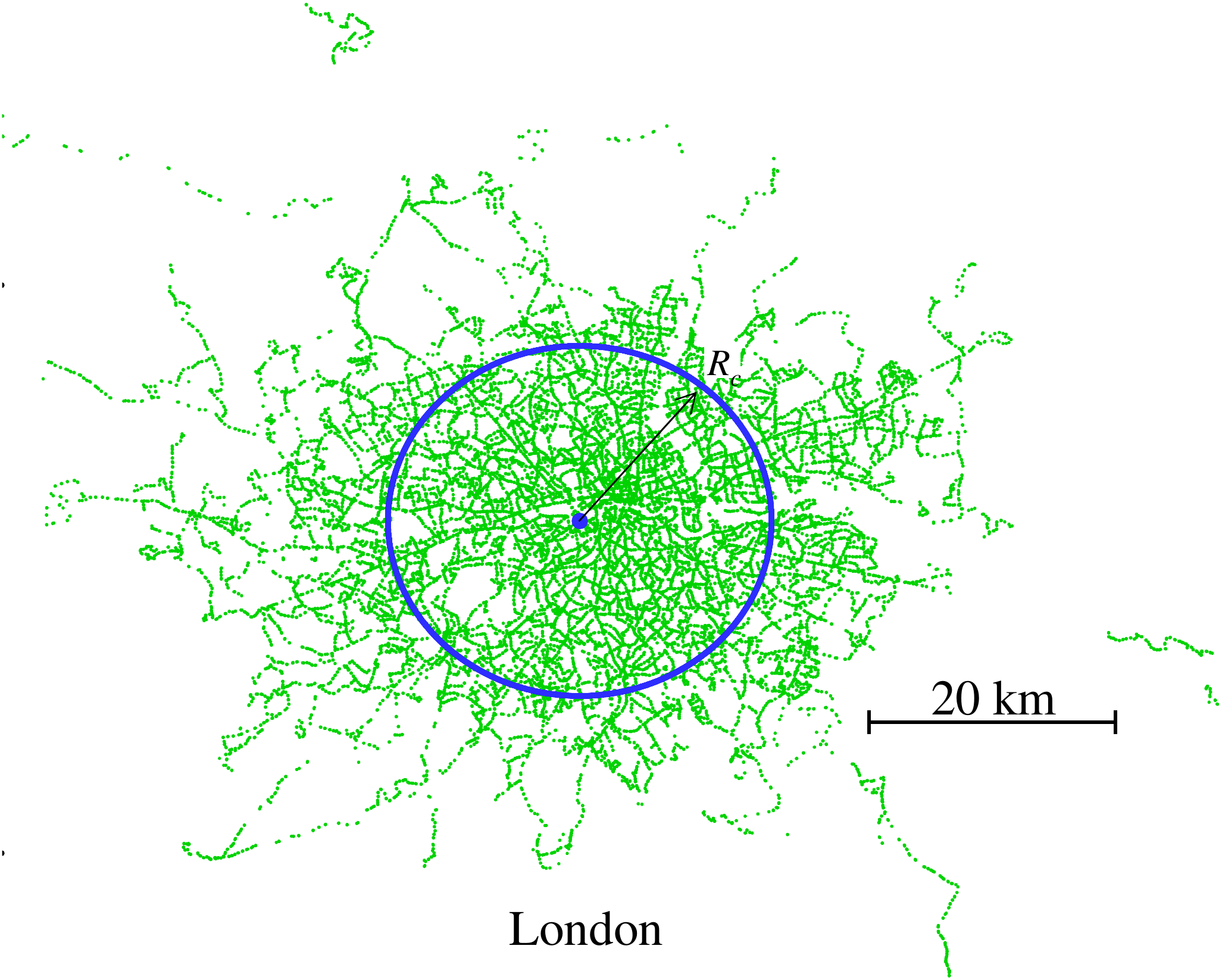}}
\caption{Example of the PTN of Greater London. The radius $R_c$
($\simeq 15.4$ km) corresponds to the transition from the compact
central area to the rarefied space with $d_f<2$.
\label{fig8}}
\end{figure}

\begin{figure}[!htb]
\centerline{
\includegraphics[width=0.9\textwidth]{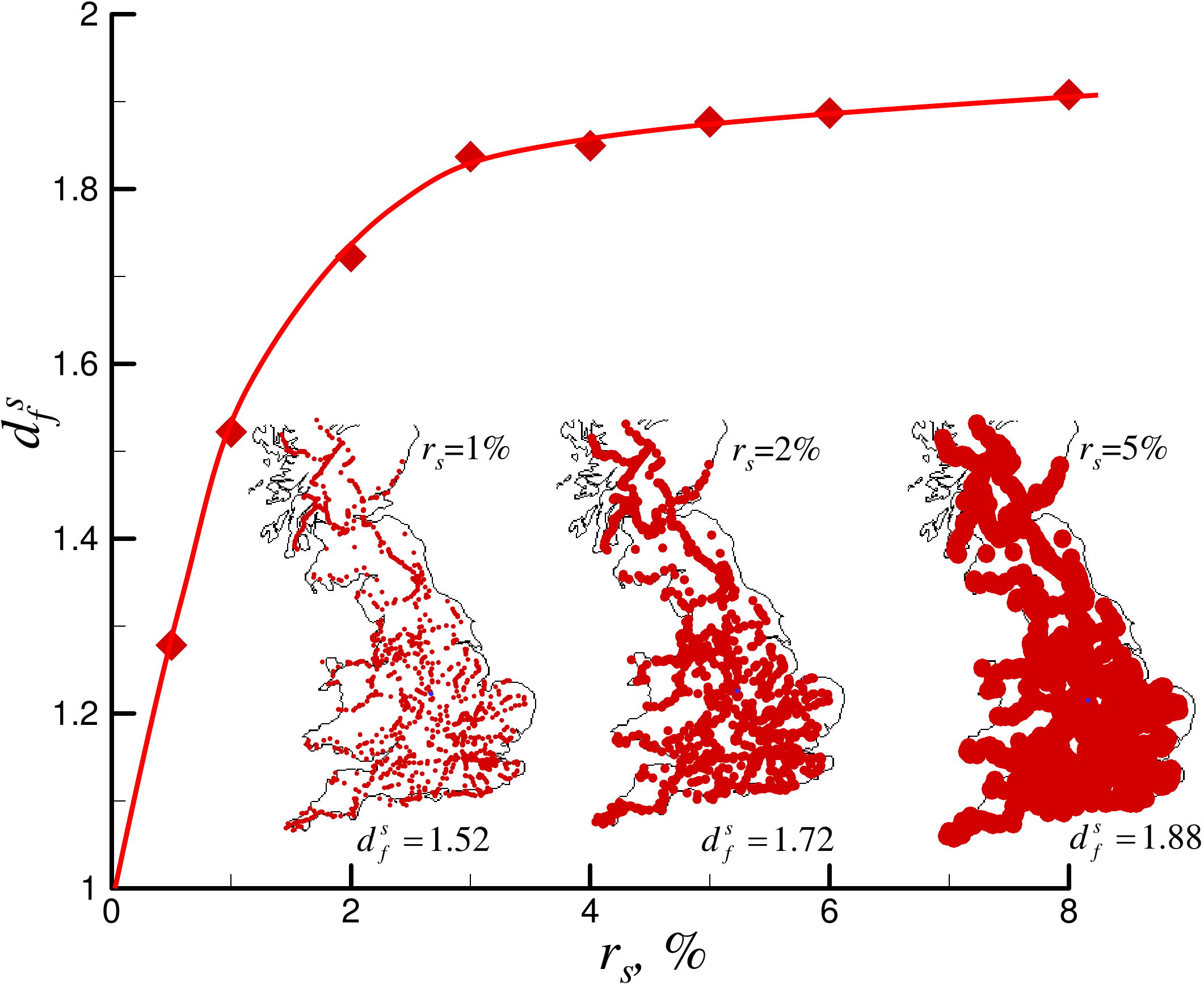}}
\caption{Fractal dimension $d_f^s$ of the GB coach network
calculated by considering a boxing method where circles of different
size $r_s$ (\% of maximum size of the object) are used to cover the
object of interest. Depending on the boxes size, one gets the value
that spans the region $d_f^s=1.28-1.91$. The inset shows examples
of the networks in geospace at different values of $r_s$.
\label{fig9}}
\end{figure}

Fig \ref{fig6} provides an example of such analysis for the GB coach and
rail networks. The outcome of similar analysis for the rest of
the networks under consideration is shown in Fig \ref{fig7}.
One can see from these values, that the fractal dimension of
national networks in the range of distances 1~km~$-$~200~km is
close to $d_f=2$ showing that these networks tend to cover uniformly
all the area they are servicing within this range. The local PTN tend 
to cover uniformly the central area with radius $R_c$ and the obvious
inhomogeneities in structure are observed at the peripheral area. In the inset of Fig \ref{fig7} the radius $R_c$ for each PTN are given. These values corresponds to the transition from the compact central area to the rarefied space with $d_f<2$, see Fig \ref{fig8} for the PTN of Greater London. This transition can be interpreted as the point at which the network ceases to provide uniform access to public transport. This transition can be interpreted as the point at which a PTN ceases to provide uniform access to commuters using public transport. 
This is an important consideration when modeling PTN. If $R_c$, is too large, this would be a waste of resources and this would 
scale rather penally according to $R = R_w - R_c$, where $R_w$ in this case represents the radius of a city. Alternatively, if $R_c$, 
is too small, the PTN would be neglecting many citizens living on the periphery of the city. Either of the two cases mentioned above are important 
to avoid when modeling PTN.
The value of $R_c$ is shown for each network. It is interesting to note that $R_c$ for Manchester is $21.8-15.4=6.4$ km larger than London. 
\\
\subsection*{Surface fractals: serviceable area of stations}
The fractal dimension can also be determined by considering a
boxing method where circles of different radii can be used to
cover the object of interest. For the GB coach network this is
illustrated in Fig \ref{fig9}. Obviously, the fractal dimensionality
$d_f^s$ calculated within this method depends on the size of 
the circles, $r_s$ used to cover the object. As one sees from
the example considered, the fractal dimensionality changes
from $d_f^s\simeq 1.28$ to $d_f^s=1.91$ as $r_s$ is increased. An interesting
interpretation of the fractal dimensionality as determined by this
method can be achieved by considering the size of a box as an area
serviced by separate public transportation stations. 
When boxes are small one ends up with the structure where $d^s_f < 2$:
effectively, the service area of all network is smaller than the
dimensionality of the geospace $d=2$. In turn, increasing the
service area of each station (i.e. increasing of the box size) leads
to an increase of $d_f^s$ finally leading to $d_f^s\simeq 2$.
Within a certain range the faster this slope grows the more evenly distributed stations
are within the network. 

\subsection*{Network Dynamics}

\begin{figure}[!htb]
\centerline{
\includegraphics[width=0.9\textwidth]{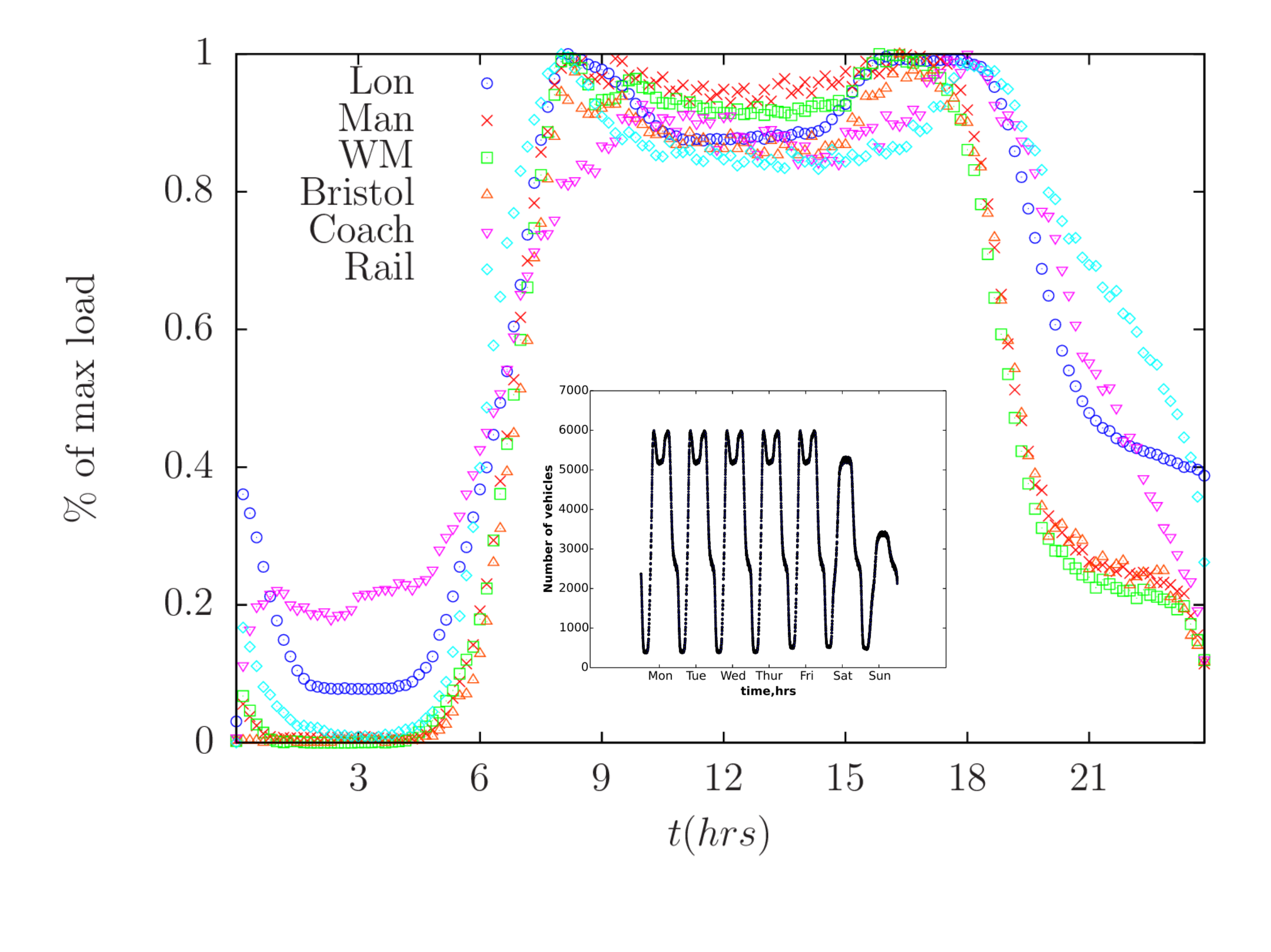}}
\caption{The normalised load for London, Manchester, West Midlands, Bristol, National coach and Rail is plotted as a 
function of time over a 24 period period (the whole of Tuesday). In the inset is the load plotted for London as a function of time for the entire week. 
\label{occ_site}}
\end{figure}

Network load measures the number of transport vehicles on the PTN at a particular point in time. Due to the fact that PTN service humans the loads on
these networks tends to follow circadian behavior. This behavior can be observed in the inset of Fig.\ref{occ_site} where the load on the London PTN is plotted for the entire week in $2010$. Here, it can be seen that for week days the loads follow a similar bimodal distribution with peaks observed during morning and evening rush hours. On the weekend, however, a unimodal distribution is observed with the loads also being lower, especially on Sunday. We can further investigate the universality of these distributions by observing other UK PTN load distributions.  

To achieve this we plot (see Fig.\ref{occ_site}) the loads for all PTNs on a particular day in the week. We can see that loads for all these PTNs 
follow similar behaviour. In particular, during the early hours of the morning load is at its lowest. This is then followed by a sharp increase in activity to a peak period. After this a slight dip in the load is observed and then the activity remains constant during the middle of the day with only minor oscillations. As the afternoon rush hour approaches another peak in the load is observed after which the load decreases sharply. However, not as sharply as in the morning increase.

Using the load $L(t)$ as a function of time the energy $E$ expended in the system can be estimated, $E=\int^{24}_{t=0} L(t)dt$. One could consider other factors 
like the speed of the transport that would effect the energy but we neglected these factors and assumed them to be equal. In turn it is tempting to consider 
how $E$ changes with an increase of city area.  

\begin{figure}[!htb]
\centering
\includegraphics[scale=0.7]{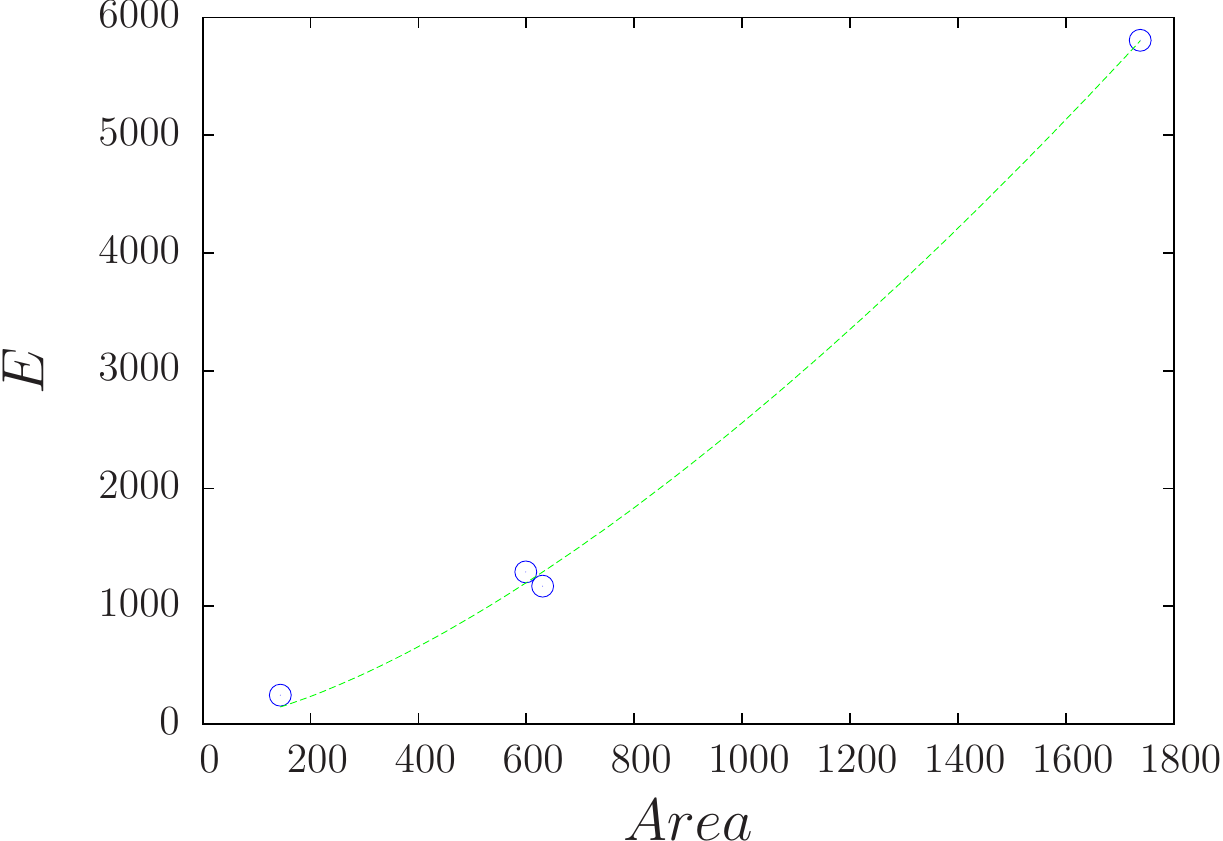}
\caption{Energy $E$/1000 as a function of geographical area in km$^2$. The blue circles indicate the respective data points and the green line is given by $E$ = Area$^{1.48}$.}
\label{load_v_area}
\end{figure}

The geographical area that cities cover can vary vastly. For example in the UK the geographical area covered by Greater London and Worcester is approximately 
$1738$ and $25$ km$^2$ respectively\footnote{The geographical area were obtained via Wikipedia} Let us get a rough estimate how the change in city size 
effects the energy consumption of public transport which is an important factor to consider when determining the costs of running transports systems. 
Let us note that the scaling effect of various city features such as number of gas stations, gas sales, length of electrical cables or road surface, populations 
wealth have been studied as a function of city size in terms of population (Bettercourt et al. 2007, 2010, 2011). An intriguing feature observed in these studies 
is that physical characteristics such as number of gas stations or length of electrical cables tend to show sub linear growth whereas wealth and other socioeconomic 
factors were found to be governed by a super linear increase. We are not aware about similar analysis of possible scaling behaviour for city characteristics as
function of area covered by the city. Although the data that is available for our analysis does not give a possibility to get a solid numerical evidence of
scaling  behaviour, we can try at least to look for a tendency in such type of dependencies. Theoretically one might expect the energy required to run a transport system would scale with area. When plotting load as a function of area (Fig.\ref{load_v_area}) we find that load increases with an exponent of $1.48$. 
This exponent has to treated with caution as only a few data points have been used to estimate it. It will be interesting to generate more data on this 
to find a better fit and hence build a better model explain how area effects the energy required to run a PTN. This is important as the energy needed to run a 
systems will be directly related to the cost of running a system. Therefore when distributing funds equally to run PTN policy makers can consider how the area 
effect running costs of PTN.          

\section{Conclusions}\label{V}

There are at least two particular features of our study that we think are worth
mentioning in the conclusions. The first, is contrary to the majority of works on
PTN where either properties in geospace or in topological space are examined,
we have completed a comprehensive analysis on both cases. The second, is the very
methodological and conceptual apparatus used in this analysis. 
Namely, we considered PTN as a graph and used concepts of complex network science to
quantify its properties.
Although the samples chosen included both local and national
public transport networks, we show that they share a lot of common properties.
\\
The main topological features of the network considered here are
summarized in table \ref{tab1}. Comparison of data for PTN with
that of a classical random graph of the same size gives significant
evidence that these networks are strongly correlated and 
assortative structures with comparatively small typical mean
shortest path length (although caution is to be made when
attributing to them small world properties). Their node degree
distributions are well described by the power law decay,
which brings about their scale free properties, at least within a
certain range of node degree values $k$.
\\
As we have emphasized above, network characteristics obtained in the course of our analysis allow for 
comparison with other PTN using a solid base of quantitative criteria. In turn, such a set of observables can be employed as KPI in aid of further developing efficient and stable PTN.
\\
As it has recently been established (Berche et al. 2012, von Ferber et al. 2012),
analysis of PTN topological features also aids in the prediction of their behaviour under
removal of their constituents. Such removal (usually called attack in the literature)
may be targeted, when the most important hubs are taken away at the first instance, or at random, when nodes are removed one by one without any preference. The last
scenario corresponds to random failures of stations that cease to operate and violate network integrity.
\\
Table \ref{tab2} shows the Molloy-Reed parameter for the GB networks that may serve as a measure of PTN stability in comparison with that for some other cities in the world. To the best of our knowledge, it has never been calculated so far for large scale transportation networks. In this sense our data for the GB national rail and coach networks provide the first example of such calculations and we wait for their comparison with their counterparts for the networks covering larger geographic space in other regions of the world.
\\

We further investigated some of the dynamical features of PTN. In particular we find their load distributions to follow similar universal behaviour. Even though these load distributions are similar in nature on a macro scale the dynamics on micro scale could be very different. One possible extension of this analysis would be to consider the temporal correlations on a micro scale. We also investigated energy as a function of geographical area finding this to scale according to the exponent $1.48$. However, with the lack of sufficient data points more data would be required to more accurately determine how the cost of running a system increases with size.   
\\
One of the corner stones of modern complexity science is generating
analogies between statistical properties of systems of
interacting agents of different nature, in particular, to study the
sensitivity of such systems to changes in their parameters (as
in the mentioned above case of targeted and random attacks), to
analyze emergent collective phenomena, to shed light on the
origin of power laws that very often govern statistics of such
systems (for a recent review see e.g. Holovatch (2017) and
references therein). These features very often are reflected in
application of concepts and methods borrowed from physics in
the out-of-physical fields. Examples from our analysis are
given by using concepts of fractal dimensions that
provide useful information on the serviceability
PTN properties in geospace. We believe that
further work in this direction will be useful both for the better
understanding of the PTN complex structure and its modeling.

\section*{Acknowledgements}
This work was supported by the EU FP7 Projects No. 612707 ``Dynamics
of and in Complex Systems" (DIONICOS), No. 612669  ``Structure
and Evolution  of Complex  Systems  with Applications in Physics and
Life Sciences" (STREVCOMS) and the National Academy of Sciences of Ukraine,
Project No. 43/18-H (ML). We would also like to thank Ralph Kenna, Petro
Sarkanych and Joeseph Yose for their useful contribution during
discussions. YuH acknowledges discussions and useful feedback from the
participants of the COST Action TD1210 'Knowescape' during the fourth
annual conference (Sofia, 22-24 February, 2017).

\newpage

\section*{Appendix A}\label{VI}
In this appendix, we provide explicit definitions for observables used to quantify different features of complex networks.

\subsection*{Mean degree $\langle k \rangle$}

For an undirected network, which is how PTN are viewed at
present in our analysis, the mean degree of the network is computed
as:
 \begin{equation}\label{1}
 \langle k \rangle = \frac{2m}{n}
 \end{equation}
 Where $n$ is the number of nodes and $m$ the number of edges. This statistic can be interpreted
 as the mean number of links of a station.

\subsection*{The Giant Connected Component (GCC)}

Strictly speaking, the giant connected component is defined as a largest connected cluster of a network
which  remains nonzero in the limit of a network of an infinite size. Here, dealing with PTN of finite size,
by the GCC we mean the largest connected part of the graph where each node has a path to every other node
in that particular section of the graph. This metric allows us to measure the connectivity of a network.

\subsection*{The mean path length $\langle \ell \rangle$}

For the connected network, the mean shortest path length $\langle
\ell \rangle$ can be defined as the average number of steps along
the shortest path for all possible pairs of nodes and gives a
measure of how closely related nodes are to each other on average.
The equation used to compute this quantity is:
\begin{equation}\label{2}
\langle \ell \rangle = \frac{2}{n(n-1)}\sum_{i \neq j} d(i,j)
\end{equation}
where $n$ is the number of nodes and $d(i,j)$ is the shortest path
between nodes $i$ and $j$. When calculating the mean $\ell$ in PTN
the GCC of the network will be used, since there is no path between
disconnected nodes. This can then be compared with the $\langle \ell
\rangle$ of a random network of the same size for which the equation
that describes how to calculate reads (Fronczak and Ho\l{}yst 2004):
 \begin{equation}\label{3}
 \langle \ell_{\rm rand} \rangle = \frac{\ln n-\alpha}{\ln(\langle k \rangle) + 0.5}
 \end{equation}
where $\alpha \approx0.5772$ is the Euler-Mascherroni constant, $n$
is the number of nodes in the network  and $\langle k \rangle$ is the mean
node degree. For the case of a weighted network considered in this paper,
defining $\langle \ell_t \rangle$ instead of adding the unit for each added
station time between the stations will be added.

\subsection*{Diameter $D$}

The diameter is the longest of all the shortest paths between two
nodes in the network. This metric is computed using the GCC only as
there is no path between disconnected segments in the graph.

\subsection*{Assortativity $r$}

Assortativity of a network is usually used to investigate
whether nodes of a similar degree tend to be linked together. This
is similar to the Pearson correlation coefficient and is calculated
as:
 \begin{equation}\label{4}
 r = \frac{A_{i,j}(k_i-E[k])(k_j-E[k])}{E[k^2-E[k]^2]}
 \end{equation}
where $A_{i,j}$ are elements of the  adjacency matrix $\hat{A}$ of
the network  ($A_{i,j}=1$ if there is  a link between nodes $i$ and
$j$  and  $A_{i,j}=0$ otherwise). $k_i$ and $k_j$ are degrees of
nodes  $i$ and $j$ respectively,  $E[k]$ is the mean node degree and
$E[k^2-E[k]^2]$ is the mean  variance of the node degree.

\subsection*{Clustering coefficient $C$}

The clustering coefficient is defined as a statistic measure of how
a network tends to cluster, i.e. if the neighbours of a given node
are also neighbours of each other. The local clustering coefficient
of node $i$ is calculated by the following equation
\begin{equation}\label{6}
 C_i = \frac{2y_i}{k_i(k_i-1)}, \hspace{2em} k_i>2,
 \end{equation}
where $k_i$ is the degree of node $i$ and $y_i$ is the number of
links between the $k_i$ nearest neighbours of the node $i$.

The mean clustering coefficient of a network is obtained as
 \begin{equation}\label{5}
 C = \frac{1}{n}\sum_{i=1}^{n} C_i
 \end{equation}
where $n$ is number of nodes in the network. It can be compared with
the mean clustering coefficient $ C_{\rm rand}$ for a random network
(Erd\"os-R\'enyi classical random graph) of the same size (Erd\"os and R\'enyi 1959; Bollobas 1985):
 \begin{equation}\label{7}
 C_{\rm rand}=\frac{\langle k \rangle}{n-1}\,\, .
 \end{equation}
Together with $ \langle \ell \rangle/ \langle \ell_{\rm rand}
\rangle$, the ratio $C/C_{\rm rand}$ can be used to decide whether a
network is of a small world type. 

\section*{Appendix B}\label{VII}

Here we show the method used to determine the function that best describes a given set of data points i.e. when determining whether data is best described by a power law $\sim ak^{-\alpha}$ or an exponential $\sim be^{-\xi/k}$. First, we determine the standard errors for the free parameters (i.e. the coefficient and exponent in the above cases) by applying a nonlinear least-squares (NLLS) Marquardt-Levenberg algorithm (Levenberg 1944).  

The standard errors are given in different scales (i.e log-lin and log-log), making them difficult to compare. Instead, we consider functions in linear space and apply integrals to determine the best fit. We find the difference in area, for each function, where standard errors give the upper and lower bound in area. From example below we give the areas calculated for a power law $A_{\text{p}}$ and exponential function $A_{\text{e}}$

\begin{equation}
A_{\text{p}} = \int\limits_{k_{\text{min}}}^{k_{\text{max}}}(a+\delta)k^{-(\alpha - \sigma)}dk
            - \int\limits_{k_{\text{min}}}^{k_{\text{max}}}(a-\delta)k^{-(\alpha + \sigma)}dk \ ,
\end{equation} 
and
\begin{equation}
A_{\text{e}} = \int\limits_{k_{\text{min}}}^{k_{\text{max}}} (b+\delta)e^{-(\xi - \sigma)/k}dk
            - \int\limits_{k_{\text{min}}}^{k_{\text{max}}}(b-\delta)e^{-(\xi + \sigma)/k}dk \ ,
\end{equation}   

where $\delta$ and $\sigma$ is the standard error for the prefactors and exponents respectively. The function that gives the least area is then considered the best fit for the given data.

\bibliographystyle{agsm}

\end{document}